\documentclass[pre, superscriptaddress, secnumarabic, tightenlines,twocolumn,showpacs,showkeys,amsmath,amssymb,nofootinbib]{revtex4}

\usepackage[mathscr]{eucal} 
\usepackage{graphicx}
\usepackage{longtable}
\newcommand{\flow}[1]{\mathcal{#1}}
\newcommand{\vect}[1]{\boldsymbol {#1}}
\newcommand{\tens}[1]{\boldsymbol{\mathsf{#1}}}
\newcommand{\vecti}{\boldsymbol{{I}}}

\newcommand{\tscale}{\theta}
\newcommand{\Vscale}{\mathcal{V}}

\DeclareMathOperator{\var}{var}
\DeclareMathOperator{\cov}{cov}

\newcommand{\ti}[1]{#1}  
\newcommand{\vol}[1]{\bf #1}  

\allowdisplaybreaks[4]
\begin{document}

\title{Steady State of a Dissipative Flow-Controlled System and the \\ Maximum Entropy Production Principle}
\author{Robert K. Niven}
\email{r.niven@adfa.edu.au}
\affiliation{School of Engineering and Information Technology, The University of New South Wales at ADFA, Canberra, ACT, 2600, Australia.}
\affiliation{Niels Bohr Institute, University of Copenhagen, Copenhagen \O, Denmark.}

\date{10 February 2009; {revisions 4 May 2009, {11 June 2009 and 4 August 2009}}}

\begin{abstract}

A theory to predict the steady state position of a dissipative, flow-controlled system, as defined by a control volume, is developed based on the Maximum Entropy (MaxEnt) principle of Jaynes, involving minimisation of a generalised free energy-like potential. The analysis provides a theoretical justification of a local, conditional form of the Maximum Entropy Production (MEP) principle, which successfully predicts the observable properties of many such systems. The analysis reveals a very different manifestation of the second law of thermodynamics in steady state flow systems, which {provides a driving force for} the formation of complex systems, including life.

\end{abstract}

\pacs{
05.70.Ln	
05.65.+b	
89.75.Fb	
}

\keywords{MaxEnt, maximum entropy production, thermodynamics, steady state, dissipative, irreversible, complex system}
\maketitle
%
%

\section{\label{Intro}Introduction} 
%
For over three decades, the Maximum Entropy Production (MEP) principle -- more precisely termed the ``maximum rate of thermodynamic entropy production'' principle -- has been found to give successful predictions for the steady state properties of a variety of dynamic, many-degree-of-freedom systems subject to flows of mass, energy, momentum, charge and/or with chemical reactions. Prominent examples include the global general circulation (atmospheric and oceanic) system of the Earth \cite{Paltridge_1975, Paltridge_1978, Paltridge_1981, Ozawa_Ohmura_1997, Paltridge_2001, Shimokawa_O_2001, Shimokawa_O_2002, Kleidon_etal_2003, Ozawa_etal_2003, Kleidon_L_book_2005, Kleidon_L_art_2005, Davis_2008, Paltridge_etal_2007} 
{and possibly other planetary bodies \cite{Lorenz_etal_2001, Goody_2007};} 
turbulent convection in a heated fluid (Rayleigh-B\'enard \cite{Benard_1901} convection) \cite{Ozawa_etal_2001},
mantle convection in the Earth \cite{Vanyo_Paltridge_1981, Lorenz_2001b, Lorenz_2002a} and moons of Jupiter and Saturn \cite{Lorenz_2001b}, 
{global Earth-biosphere water and nutrient cycles \cite{Kleidon_2004, Kleidon_L_book_2005, Kleidon_S_2008}},
{vegetation spatial distributions \cite{Kleidon_F_L_2007},}
biochemical metabolism \cite{Meysman_B_2007} 
and 
ecosystem operation \cite{Bruers_M_2007}.  
The Earth climate models include quite accurate predictions of the mean latitudinal air temperature, fractional cloud cover, meridional heat flux \cite{Paltridge_1975}, mean vertical air temperature profile, vertical heat flux \cite{Ozawa_Ohmura_1997} and historical latitudinal air temperature gradients over decadal and glacial-interglacial time scales \cite{Davis_2008}.  The MEP principle has also been invoked for the explanation of 
shear turbulence (Couette flow) \cite{Ozawa_etal_2001}, 
incompressible and compressible turbulent fluid flow \cite{Paulus_G_2004}, 
currents in electrical circuits \cite{Zupanovic_etal_2004, Botric_etal_2005, Christen_2006, Bruers_etal_2007a}, 
plasma formation \cite{Christen_2007a} and structure \cite{Yoshida_M_2008},
crystal growth \cite{Martyushev_A_2003},
chemical cycle kinetics \cite{Zupanovic_J_2004}, 
diffusion in non-uniform solids \cite{Christen_2007b}, 
photosynthesis mechanism \cite{Juretic_Z_2003}, 
biomolecular motors \cite{Dewar_etal_2006}
and
economic activity \cite{Jenkins_2005}; several detailed reviews are available \cite{Ozawa_etal_2003, Kleidon_L_book_2005, Martyushev_S_2006, Bruers_2007c}.  {The MEP principle has even been invoked as a technical basis for the Gaia hypothesis \cite{Lovelock_1988, Kleidon_2004, Kleidon_L_art_2005, Karnani_A_2009}.} 
The apparent successes of the MEP principle -- which can exclude most details of the dynamics -- contradict the prevailing paradigm of developing ever-more-complicated dynamic models of complex systems, for example the general circulation models (GCMs) of the Earth climate system. This suggests that in a large class of steady state systems, the dynamics adjust themselves to achieve a state of MEP, rather than the entropy production being a consequence of the dynamics.  This implies the action of an as-yet unrecognised physical principle applicable to flow systems -- beyond the domain of present-day thermodynamics -- which provides the driving force in the system, and so can be used as a ``short-cut'' in system modelling. 

In recent years, several workers have attempted to justify the MEP principle on theoretical grounds. Dewar \cite{Dewar_2003, Dewar_2005} analysed a time-variant, non-equilibrium system in terms of its available paths in parameter space; the transient and steady state positions are inferred by the maximum entropy (MaxEnt) method of Jaynes \cite{Jaynes_1957, Jaynes_1957b, Jaynes_1963, Tribus_1961a, Tribus_1961b, Kapur_K_1992, Jaynes_2003}, using an entropy defined on the set of paths. The analysis has received some criticisms \cite{Bruers_2007d, Grinstein_L_2007}, to suggest it might apply only in the near-equilibrium linear (Onsager) regime, in which fluxes are linearly proportional to their driving gradients \cite{Onsager_1931a, Onsager_1931b}. Attard \cite{Attard_2006a, Attard_2006b} also gives an analysis based on an entropy defined on transition probabilities, but cast in the terminology of traditional statistical mechanics (although he denies the MEP principle in favour of an alternative). {Beretta \cite{Beretta_2001, Gyft_Beretta_2005, Beretta_2006, Beretta_2008} examines a steepest entropy ascent principle for transitions between states, based on a quantum thermodynamics formulation.} 
Other arguments for MEP have been given by \v{Z}upanovi\'c and co-workers \cite{Zupanovic_etal_2006}, 
{based on the ``most probable'' increase in entropy during a fluctuation} 
(also criticised \cite{Bruers_2007d}), and by Martyushev \cite{Martyushev_2007}, {based on a frame of reference (relative velocity) argument}.  Of course, this is a well-trodden field, with many historical antecedents to the MEP principle (see the fascinating review by Martyushev and Seleznev \cite{Martyushev_S_2006}). If the above objections can be overcome, the path-based analyses hold the promise of predicting the behaviour of time-variant systems (transport phenomena).  However, they seem unnecessarily complicated for the task of predicting the steady state position, when (as will be shown) a more direct method is available. 

The aim of this work is to outline a theory to {\it directly} predict the steady state position of a flow-controlled system, based on Jaynes' MaxEnt method \cite{Jaynes_1957, Jaynes_1957b, Jaynes_1963, Tribus_1961a, Tribus_1961b, Kapur_K_1992, Jaynes_2003}. The analysis provides a theoretical {derivation of a local, conditional form of} the MEP principle for steady state thermodynamic systems of any type (e.g.\ heat flow, fluid flow, diffusive flow, electrical flow, chemical and process engineering, biological and human systems). The approach taken here differs from previous analyses \cite{Dewar_2003, Dewar_2005, Attard_2006a, Attard_2006b, Zupanovic_etal_2006} in that it considers flux rather than path probabilities, to directly obtain the steady state position. It also employs the local equilibrium assumption commonly used in engineering control volume analysis, but does not appear to be restricted to the linear transport regime; {indeed, the linear regime emerges as a first-order approximation to the analysis}.

This work is set out as follows. In \S\ref{Jaynes}, the principles of Jaynes' generic theory are outlined, leading in \S\ref{systems} to a discussion of different types of physical systems, which influences the manner in which Jaynes' method can be applied. Equilibrium systems are first examined, to demonstrate the universality of Jaynes' method and to highlight the (easily overlooked) equilibrium analogue of the MEP principle.  In \S\ref{steady_state}, a theory is developed to predict the steady state position of a flow-controlled thermodynamic system; by comparison to a traditional control volume analysis, this is found to give a local, conditional form of the MEP principle. 
The implications of the analysis for the formation of complex systems, including life, are discussed in \S\ref{Discussion}.
%
\section{\label{Jaynes}Jaynes' Generic Theory} 
%
We first summarise Jaynes' generic formulation of statistical mechanics \cite{Jaynes_1957, Jaynes_1957b, Jaynes_1963, Tribus_1961a, Tribus_1961b, Kapur_K_1992, Jaynes_2003}, based on the {minimum divergence}, {maximum relative entropy} or {minimum cross-entropy} (MinXEnt) principle, within which the {maximum entropy} (MaxEnt) principle can be considered a special case. Consider a system of $N$ discrete, distinguishable entities distributed amongst $s$ distinguishable categories or ``states'' within a system, with constant source (``prior'') probabilities $q_i$ for the filling of each state $i=1,...,s$ (the states $i$ may be multivariate, e.g.\ $\{i,j,...\}$).  From information-theoretic considerations \cite{Shannon_1948, Shore_J_1980} and/or by combinatorial arguments \cite{Boltzmann_1877, Planck_1901, Vincze_1972, Grendar_G_2001, Niven_CIT, Niven_MaxEnt07}, it can be shown that in the asymptotic limit $N \to \infty$ (the Stirling approximation \cite{Stirling_1730} or Sanov's theorem \cite{Sanov_1957}),  the ``least informative'' or ``most probable'' distribution of the system is obtained by minimising the Kullback-Leibler cross-entropy function \cite{Kullback_L_1951, Kullback_1959} (the negative of the relative entropy function \cite{Jaynes_1963})\footnote{Use of the continuous form of \eqref{eq:KL}, or the von Neumann entropy based on density matrices, requires different mathematics but gives essentially the same macroscopic results \eqref{eq:pstar2_i}-\eqref{eq:dDstar} \cite{Jaynes_1957b, Jaynes_1963}.}:
\begin{equation}
D = \sum\limits_{i = 1}^s {p_i \ln \frac{{p_i }}{{q_i }}} 
\label{eq:KL}
\end{equation}
subject to the natural and moment constraints on the system, respectively:
\begin{gather}
\sum\limits_{i = 1}^s {p_i }= 1,
\label{eq:C0} 
\\
\sum\limits_{i = 1}^s {p_i f_{ri} }= \langle {f_r } \rangle, \quad r = 1,...,R, 
\label{eq:Cr}
\end{gather}
where $p_i$ is the probability of the $i$th category, $f_{ri}$ is the value of property $f_r$ for the $i$th category and $\langle {f_r} \rangle$ is the mathematical expectation of $f_{ri}$ (generally, each $f_r$ is a ``conserved quantity'').  Applying Lagrange's method of undetermined multipliers to (\ref{eq:KL})-(\ref{eq:Cr}) gives the stationary or most probable distribution of the system:
\begin{gather}
\begin{split}
p_i^* &=  q_i e^ {  - \lambda_0^*   - \sum\limits_{r = 1}^R  {\lambda_r  f_{ri} }    } =  (Z_q^*)^{-1} { q_i e^ {  - \sum\limits_{r = 1}^R  \lambda_r  f_{ri}   } }  , 
\\
Z_q^* &=  e^{\lambda_0^*} = \sum\limits_{i=1}^s {q_i e^ {  - \sum\limits_{r = 1}^R  \lambda_r  f_{ri}   }} 
\end{split}
\label{eq:pstar2_i}
\end{gather}
where $\lambda_r$ is the Lagrangian multiplier for the $r$th constraint, $\lambda_0^*$ is the ``Massieu function'' \cite{Massieu_1869}, $Z_q^*$ is the generalised partition function and ``$^*$'' denotes a quantity at the stationary position. Here ``stationary'' implies a time-invariant distribution \eqref{eq:pstar2_i}, with no additional physical interpretation.  Subsequent generic analyses pioneered by Jaynes \cite{Jaynes_1957, Jaynes_1957b, Jaynes_1963, Tribus_1961a, Tribus_1961b, Kapur_K_1992, Jaynes_2003, Niven_CIT} give the relations:
\begin{gather}
{ D^*  = - \lambda_0^*  -  \sum\limits_{r = 1}^R \lambda_r \langle {f_r} \rangle }
\label{eq:MinXEnt}
\\
\frac{\partial  \lambda_0^* }{\partial \lambda_r}  = - \langle {f_r} \rangle 
\label{eq:diff1}
\\
\frac{\partial^2 \lambda_0^*}{\partial \lambda_r^2}  = \langle {{f_r}^2} \rangle- {\langle {f_r} \rangle}^2   = - \frac {\partial \langle {f_r} \rangle}{\partial \lambda_r}  \label{eq:diff2} 
\\
\begin{split}
\frac{\partial^2 \lambda_0^*}{\partial \lambda_m \partial \lambda_r}   = \langle {{f_r}{f_m}} \rangle- {\langle {f_r} \rangle}{\langle {f_m} \rangle}   
= - \frac {\partial \langle {f_r} \rangle}{\partial \lambda_m}  = - \frac {\partial \langle {f_m} \rangle}{\partial \lambda_r}
\label{eq:diff2mixed}
\end{split}
\\
{dD^* = - \sum\limits_{r = 1}^R {\lambda _r \Bigl( d\langle {f_r } \rangle  - \delta W_r \Bigr)}  = - \sum\limits_{r = 1}^R {\lambda _r \delta Q_r } }
\label{eq:dDstar}
\end{gather}
where $\delta W_r = \langle {df_r } \rangle = \sum\nolimits_{i=1}^s  p_i^* d f_{ri}$ and $\delta Q_r = \sum\nolimits_{i=1}^s  d p_i^* f_{ri}$ can be identified, respectively,  as the increments of ``generalised work'' on the system 
and ``generalised heat'' added to the system during a change in the $r$th constraint $d \langle {f_r } \rangle$, in which $\delta x$ denotes a path-dependent differential. (A {\it path} is here defined as a specified trajectory in macroscopic coordinates, e.g.\ constant constraints $\{\langle f_r \rangle\}$, constant multipliers $\{\lambda_r\}$ or some mixed form.) To avoid double counting, the $\delta W_r$ terms represent generalised work processes {\it in addition to} those {introduced by other constraints}{\footnote{{This point is further explained in \S\ref{equilsys} and footnote \ref{foot:PV}, by reference to an example equilibrium system.}}}.

For equal $q_i=s^{-1}$, the foregoing analysis reduces to that of maximising the Shannon entropy \cite{Shannon_1948}:
\begin{equation}
\mathfrak{H} =   - \sum\limits_{i = 1}^s {p_i \ln p_i } 
\label{eq:Shannon}
\end{equation}
subject to the same constraints (\ref{eq:C0})-(\ref{eq:Cr}).  This gives the same relations \eqref{eq:pstar2_i}-\eqref{eq:dDstar}, except that $D^*$ is replaced by $-\mathfrak{H}^*$ and the $q_i$ cancel from \eqref{eq:pstar2_i}.  

It is important to note that relations \eqref{eq:pstar2_i}-\eqref{eq:dDstar} only concern the stationary position, i.e.\ they apply ``on the manifold of stationary positions''. Transitions between stationary positions are thus assumed to be {\it quasistatic}, i.e.\ they are idealised as taking place only between stationary positions, without any intermediate non-stationary positions \cite{Callen_1985}. Comparing \eqref{eq:dDstar} to the derivative of \eqref{eq:MinXEnt} gives the differential:
\begin{align}
\begin{split}
d \phi^* &= - d \lambda_0^*  = - d \ln Z_q^* =  \sum\limits_{r = 1}^R {\lambda _r  \delta W_r }  +  \sum\limits_{r = 1}^R d \lambda_r \langle {f_r} \rangle
\\
&=dD^*  + \sum\limits_{r = 1}^R \lambda_r d \langle {f_r} \rangle +\sum\limits_{r = 1}^R d \lambda_r \langle {f_r} \rangle
\end{split}
\label{eq:Massieu}
\raisetag{20pt}
\end{align}
$d\phi^*$ therefore encompasses ``all possible changes'' in the stationary position, due to changes in the constraints, multipliers or cross-entropy (generalised heat input). For constant multipliers, $d \phi^*|_{\{\lambda_r\}}$ is equal to the weighted change in generalised work on the system; its integrated form $\phi^*$ can therefore be regarded as a generalised, dimensionless free energy-like function or {\it generalised potential} \cite{Jaynes_1957, Jaynes_1957b, Jaynes_1963, Tribus_1961a, Tribus_1961b, Jaynes_2003}. 

We now come to the central tenet of this analysis, the importance of which will become clear in \S\ref{systems}-\S\ref{steady_state}.  If a dynamic system described by \eqref{eq:pstar2_i}-\eqref{eq:dDstar} and \eqref{eq:Massieu} contains mechanism(s) by which its generalised potential $\phi^*$ can be dissipated (converted to an unrecoverable form), by processes not represented in the constraints, then by successive increments $d \phi^*<0$ it will converge to a stationary position of no further extractable generalised work $d \phi^*=0$, equivalent to the minimum position $\phi^*=\phi^*_{\min}$.  We here confine the discussion to systems with {\it reproducible} dissipative phenomena, whereupon $\phi^*_{\min}$ will be predictable (i.e.\ $\phi^*$ is a state function).  If the incremental changes are restricted in some manner -- e.g.\ the system is confined to a constant-constraint $\{\langle f_r \rangle\}$ or a constant-multiplier $\{\lambda_r\}$ path -- then $\phi^*_{\min}$ will be the local (path-dependent) minimum. If there is no such path restriction, $\phi^*_{\min}$ will constitute the global minimum. 

It is emphasised that the above analysis (\ref{eq:KL})-\eqref{eq:Massieu} is generic, and applies to {\it any} probabilistic system which can be analysed by the Kullback-Leibler cross-entropy (\ref{eq:KL}) or Shannon entropy (\ref{eq:Shannon}). Although its primary application has been to equilibrium thermodynamics, the analysis has far broader power of application \cite{Jaynes_1957, Jaynes_1957b, Jaynes_1963, Tribus_1961a, Tribus_1961b, Kapur_K_1992, Jaynes_2003}.  For this reason, the symbols used above are generic, and should not be interpreted in terms of particular thermodynamic quantities except when so stated (e.g.\ the generic entropy $\mathfrak{H}$ should not be confused with the thermodynamic entropy $S$).
%
\section{\label{systems}Types of Systems} 
\subsection{\label{equilsys}Quantity-Constrained (Equilibrium) Systems}

\begin{figure}[h]
\begin{center}
\setlength{\unitlength}{0.6pt}
  \begin{picture}(350,480)
   \put(50,0){\includegraphics[width=55mm]{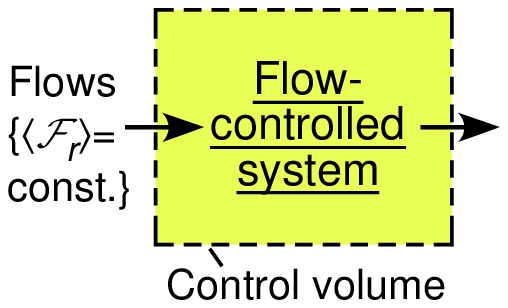} }
   \put(0,0){\small (c)}
   \put(50,160){\includegraphics[width=55mm]{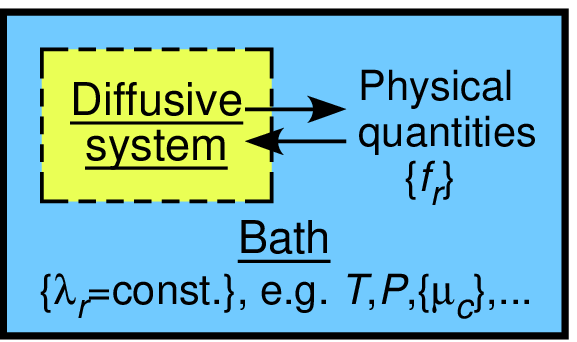} }
   \put(0,160){\small (b)}
   \put(50,320){\includegraphics[width=55mm]{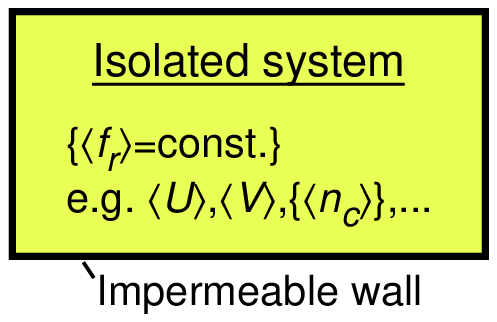} }
   \put(0,320){\small (a)}
  \end{picture}
\end{center}
\caption{{(Color online)} Different types of physical systems: (a) isolated, (b) diffusive and (c) flow-controlled (control volume).}
\label{fig:systems}
\end{figure}

To understand the application of MinXEnt, we now consider several types of probabilistic systems composed of discrete entities, as illustrated in Figure \ref{fig:systems}. Historically, Jaynes' method (and its predecessor, traditional statistical mechanics) has been applied to (a) {\it isolated} (microcanonical) systems (Figure \ref{fig:systems}a), held in a state of constant mean contents $\langle {f_r } \rangle $ of various physical quantities $f_r$ by isolation from the rest of the universe; and (b) various types of {\it open} or {\it diffusive} (e.g.\ canonical, grand canonical) systems (Figure \ref{fig:systems}b), open to the diffusion of various quantities $f_r$ but with no directed flows, and in contact with infinite {\it generalised baths} of constant corresponding $\lambda_r$ \cite{Jaynes_1957}. Clearly, such representations are convenient models 
to enable the construction of thermodynamic relationships:\ no system can really be isolated from the rest of the universe, whilst the mechanisms used to maintain the generalised baths are not usually explained. Notwithstanding this criticism, in either case the ``least informative'' or ``most probable'' position of the system -- the {\it equilibrium position} -- can be calculated by the application of MinXEnt (or, for equal $q_i$, by MaxEnt) subject to the natural and moment constraints \cite{Jaynes_1957}.  Jaynes' generic formulation (\S\ref{Jaynes}) can then be applied to the analysis of such systems.

As an example, consider an ``open'' or ``heterogeneous'' thermodynamic system for which the ``entities'' are interpreted either as discrete particles (atoms, ions, molecules, etc) or, in the Gibbs-Einstein representation, as duplicates of the entire system \cite{Gibbs_1902, Einstein_1902, Einstein_1903}. Each entity has the choice of its (quantised) internal energy $U_i$, $i=1,...,s$; volume element $V_j$, $j=1,...,t$; and moles of particles $n_{N_c}$ of each chemical species $c=1,...,C$, for which the number of particles $N_c$ can range between zero and (effectively) infinity.  The system is constrained by its natural constraint (\ref{eq:C0}), mean internal energy $\langle {U} \rangle$, mean volume $\langle V \rangle$ and mean moles of particles $\langle {n_c} \rangle$ of each type $c$. Assuming, from the principle of insufficient reason \cite{Jaynes_2003}, that each multivariate state $\{i,j,\{N_c\}\}$ is equiprobable, we maximise the Shannon entropy:
\begin{align}
\mathfrak{H}_{eq} &=   -  \sum\limits_{i=1}^s   \sum\limits_{j=1}^t  \sum\limits_{N_1=0}^\infty  \sum\limits_{N_2=0}^\infty  ... \sum\limits_{N_C=0}^\infty   {p_{i,j,\{N_c\}} \ln p_{i,j,\{N_c\}} } ,
\label{eq:H_eq}
\end{align}
subject to the constraints, giving the equilibrium position (\ref{eq:pstar2_i}):
\begin{gather}
\begin{split}
p_{i,j,\{N_c\}}^* &= \dfrac {\alpha_{i,j,\{N_c\}}^*  }  {\Xi^*}, \\
\alpha_{i,j,\{N_c\}}^* &= \exp \bigl( {  - \lambda_{U} U_i  -\lambda_V V_j - \sum\limits_{c = 1}^C  \lambda_c  n_{N_c}   } \bigr), \\
\Xi^* =  e^{\lambda_0^*} &=       \sum\limits_{i=1}^s   \sum\limits_{j=1}^t  \sum\limits_{N_1=0}^\infty  \sum\limits_{N_2=0}^\infty  ... \sum\limits_{N_C=0}^\infty  { \alpha_{i,j,\{N_c\}}^* }.
\end{split}
\label{eq:pstar2_i_thermo} 
\end{gather}
From our existing (historical) knowledge of thermodynamics (e.g.\ by monotonicity arguments \cite{Kapur_K_1992} or from the zeroth law of thermodynamics \cite{Tribus_1961b, Kapur_K_1992}), we can identify the Lagrangian multipliers $\lambda_{U}=1/kT$, $\lambda_V = P/kT$ and $\lambda_c=-\mu_c /kT$ as functions of the Boltzmann constant $k$, absolute temperature $T$, absolute pressure $P$ and molar chemical potential $\mu_c$ of each species $c$, whence:
\begin{gather}
\alpha_{i,j,\{N_c\}}^* =  \exp \Bigl( {  -  \frac{U_i}{kT} - \frac{P V_j}{kT}  + \sum\limits_{c = 1}^C  \frac{ \mu_c n_{Nc}}{kT } } \Bigr),
\label{eq:pstar3_i_thermo} 
\end{gather}
and $\Xi^*$ is the grand partition function. Recognising the thermodynamic entropy as $S=k\mathfrak{H}_{eq}$, (\ref{eq:MinXEnt}) and (\ref{eq:dDstar}) give:
\begin{gather}
S^*   = k \ln \Xi^*  +  \frac{ \langle {U} \rangle}{T} +  \frac{ P \langle {V} \rangle}{T}  - \sum\limits_{c=1}^C \frac{ \mu_c \langle {n_c} \rangle }{T} 
\label{eq:Sstar}
\\
{
\begin{split}
&dS^*  = \frac{ \delta Q_U}{T} + \frac{ P \delta Q_V}{T} 
- \sum\limits_{c=1}^C {  \frac{ \mu_c \delta Q_{n_c}}{T} } 
\\
&=  \frac{1}{T} \bigl(d\langle {U} \rangle - \delta W_U \bigr) 
+  \frac{P}{T} \bigl( d\langle {V} \rangle - \delta W_V \bigr)
\\ &\qquad 
- \sum\limits_{c=1}^C   \frac{\mu_c}{T}  \bigl( d \langle {n_c} \rangle - \delta W_{n_c} \bigr)
\\
&=  \frac{ d\langle {U} \rangle}{T}   
+  \frac{ P d\langle {V} \rangle}{T}  - \frac{\sum \delta W}{T} 
- \sum\limits_{c=1}^C   \frac{\mu_c d \langle {n_c} \rangle}{T}  
- \sum\limits_{c=1}^C   \frac{\mu_c \delta W_{n_c}}{T}  
\end{split}
}
\label{eq:dSstar}
\raisetag{78pt}
\end{gather}
{
where $\delta Q_U$, $\delta Q_V$ and $\delta Q_{n_c}$ are the increments in generalised heats, and  $\delta W_U$, $\delta W_V$ and $\delta W_{n_c}$ are the increments in generalised work, associated respectively with constraints $\langle U \rangle$, $\langle V \rangle$ and $\langle n_c \rangle$. In the first line of \eqref{eq:dSstar}, the first two terms can be amalgamated into the (actual) heat term $\delta Q/T$ used in thermodynamics; for systems with invariable particle numbers, $\delta Q_{n_c}=0$ and \eqref{eq:dSstar} reduces to the Clausius \cite{Clausius_1865} equality $dS^*=\delta Q/T$. Similarly, in the last line of \eqref{eq:dSstar}, the $\delta W_U$ and $P \delta W_V$ terms are amalgamated into the combined (actual) work $\sum \delta W$; the latter therefore represents the total (actual) work on the system, in addition to the $P d\langle {V} \rangle$ and ${\mu_c d \langle {n_c} \rangle}$ work\footnote{{\label{foot:PV}As mentioned in \S\ref{Jaynes}, the $\delta W_r$ refer to work terms in addition to those introduced by the other constraints. Thus in the above equilibrium system, the $P d \langle V \rangle$ work cannot be introduced within the work term $\delta W_U$ associated with $\langle U \rangle$, since it is already included by virtue of the $\langle V \rangle$ constraint. Similarly, $\delta W_U$ cannot include the chemical potential work terms $\mu_c d \langle n_c \rangle$, since these are included by the $\langle n_c \rangle$ constraints.  This consideration is quite general: the constraints $\langle f_r \rangle$ must be linearly independent, but need not be orthogonal; the weighted generalised work terms $\lambda_r \delta W_r$ therefore cannot include processes represented by $\lambda_m d \langle f_m \rangle$, $m \ne r$.}}. 
For this study, we ignore relativistic or other changes in the ``mass levels'', and set $\sum \delta W_{n_c}=0$.} It is again emphasised that \eqref{eq:pstar2_i_thermo}-\eqref{eq:dSstar} apply only to quasistatic transitions, in this case ``on the manifold of equilibrium positions''.

Eqs.\ \eqref{eq:Massieu} and \eqref{eq:Sstar}-\eqref{eq:dSstar} give the change in the (actual) free energy function, in energy units:
\begin{align}
\begin{split}
dJ^* &= k T d\phi_{eq}^*= - k T d\ln \Xi^*  \\
&= - T dS^*   +  { d \langle {U} \rangle}  +  { P d \langle {V} \rangle }  - \sum\limits_{c=1}^C {  { \mu_c  d \langle {n_c} \rangle}  }   \\
& \quad +  T \biggl\{ d \Bigl( \frac{ 1}{T} \Bigr) \langle {U} \rangle +  d \Bigl( \frac{ P }{T}  \Bigr)  \langle {V} \rangle - \sum\limits_{c=1}^C {  d \Bigl( \frac{ \mu_c  }{T} \Bigr) \langle {n_c} \rangle } \biggr\}
\end{split}
\label{eq:dJstar}
\raisetag{78pt}
\end{align}
in which, from \eqref{eq:dSstar}, the first four terms on the right correspond to the total work $\sum\delta W$. Note that -- as per Gibbs \cite{Gibbs_1873b, Gibbs_1875} -- we could restrict \eqref{eq:dJstar} to these first four terms, by imposing the Gibbs-Duhem relation 
$\sum\nolimits_{r=1}^R d\lambda_r \langle f_r \rangle=0$.
We here wish, however, to consider all possible changes in $J^*$, allowing both Gibbs-Duhem paths and non-Gibbs-Duhem paths. Integration of \eqref{eq:dJstar} from a zero reference state then yields\footnote{Strictly speaking, we must integrate $k \, d\phi_{eq}^*$ and multiply by $T$, an awkwardness produced by the conversion of $\ln\Xi^*$ to energy rather than entropy units (see Appendix \ref{Apx_termin}). Alternatively, integration of $dJ^*|_{T,P,\{\mu_c\}}$ directly yields \eqref{eq:J} \cite{Gibbs_1875}.}:
\begin{equation} 
\begin{split}
J^*&=kT \phi^*_{eq} =-kT\ln\Xi^* 
\\
&= - T S^*  +  \langle {U} \rangle +  { P \langle {V} \rangle}  - \sum\limits_{c=1}^C {  { \mu_c \langle {n_c} \rangle } }
\end{split}
\label{eq:J}
\end{equation}
Although less well known, $J^*$ was reported by Gibbs \cite{Gibbs_1875}; in modified form, it forms the basis of the available energy, essergy and exergy concepts \cite{Gibbs_1875, Keenan_1941, Keenan_1951, Rant_1956, Gaggioli_1962, Evans_1969, 
Moran_S_2006}. For constant composition, $J^*$ reduces to the Gibbs free energy $G^*= - T S^* + {  \langle {U} \rangle} +  { P  \langle {V} \rangle}$ plus a constant, whilst for constant composition and volume, it gives the Helmholtz free energy $F^*= - T S^* + { \langle {U} \rangle}$ plus a constant.

We first examine \eqref{eq:dJstar}-\eqref{eq:J} from the original perspective of Gibbs [\onlinecite{Gibbs_1873b, Gibbs_1875}; see also \onlinecite{Callen_1985, Gaggioli_etal_2002a, Gaggioli_etal_2002b}].  Following the standard treatment, we consider an open system surrounded by an intensive variable bath (Figure \ref{fig:systems}b), with the double system-bath isolated from the rest of the universe; a single isolated system (Figure \ref{fig:systems}a) can then be considered as a special case. From \eqref{eq:dSstar}-\eqref{eq:dJstar}, $dJ^*|_{T,P,\{\mu_c\}}=\sum\delta W$ gives the total change in work on the system along a Gibbs-Duhem path, due to processes not represented in the constraints.  Thus $dJ^*|_{T,P,\{\mu_c\}}<0$ indicates spontaneous change (work extraction) and $dJ^*|_{T,P,\{\mu_c\}}>0$ a forced change (work input). If the system contains some reproducible dissipative process(es) by which work can be expended (e.g.\ friction \cite{Tribus_1961a, Tribus_1961b} or chemical reactions \cite{Prigogine_1967, Kondepudi_P_1998}), it will move towards the equilibrium defined by $dJ^*|_{T,P,\{\mu_c\}}=0$, or in other words, to the minimum free energy position $J^*=J^*_{min}$. This was implicitly understood by Gibbs \cite{Gibbs_1873b}, who refers to the equilibrium surface (the ``fundamental relation'' or Euler surface) as the ``surface of dissipated energy''. If no work is extracted in useful form, each increment of lost work is given by \cite{Planck_1922, Planck_1932, Fermi_1936}:
\begin{equation}
dJ^*|_{T,P,\{\mu_c\}} = -T dS^*|_{T,P,\{\mu_c\}} - T \delta \sigma|_{T,P,\{\mu_c\}} \le 0
\label{eq:dJstardissip1}
\end{equation}
where $dS^*|_{T,P,\{\mu_c\}}$ and $\delta \sigma|_{T,P,\{\mu_c\}}$ respectively give the change in entropy within and outside the system. 
Eq.\ \eqref{eq:dJstardissip1} represents the net irreversible increase in entropy in the double system-bath (converted to energy units and with change of sign), reflecting the interplay between irreversible change within the system, and the exporting of irreversible change to the bath by the transfer of generalised heats \citep[c.f.][]{Planck_1922, Planck_1932, Fermi_1936, Strong_H_1970, Craig_1988}.  The second term $\delta \sigma|_{T,P,\{\mu_c\}}$ can be labelled the {\it entropy produced}\footnote{It is misleading to call $\sigma$ the {\it entropy production}. In English, ``{production}'' implies an on-going phenomenon, hence a rate process (e.g.\ a nation's steel production). This can only apply to a flow system. Confusingly, the symbol $\sigma$ is used interchangeably for the amount of thermodynamic entropy produced by a system, its rate of production by a system or the rate per unit volume; these are labelled $\sigma$, $\dot{\sigma}$ and $\hat{\dot{\sigma}}$ respectively here.} 
by the system, since it materialises outside rather than inside the system. If the incremental losses are unrecoverable (i.e.\ $dS^*|_{T,P,\{\mu_c\}} \ge 0$ and $\delta \sigma|_{T,P,\{\mu_c\}} \ge 0$), the equilibrium position $J^*=J^*_{min}$ will correspond to the position of maximum entropy produced, $\sigma=\sigma_{\max}$ \citep[c.f.][]{Kondepudi_P_1998}. 

We now go a step farther than Gibbs, to consider systems which can deviate from a Gibbs-Duhem path during dissipation. We again only consider systems with reproducible processes. Here it is preferable to use the potential $d\psi^*=k d\phi^*_{eq}=dJ^*/T$ (the grand potential form of the negative Massieu function \cite{Massieu_1869} or negative Planck potential \cite{Planck_1922, Planck_1932, Schrodinger_1946, Guggenheim_1967, Strong_H_1970, Craig_1988}; see Appendix \ref{Apx_termin}) rather than $dJ^*$, to more effectively handle changes in temperature.  From \eqref{eq:dJstar}, each increment $d\psi^*<0$ can be divided into two parts: (i) the part $d\psi^*|_{S^*, \langle U \rangle, \langle V \rangle,\{\langle n_c \rangle\}}<0$ which is expended against changes in the intensive variables, causing an irreversible loss of ``ability to do generalised work'', manifested as an increase in entropy within or outside the system, and (ii) the part $d\psi^*|_{T,P,\{\mu_c\}}<0$, which -- as in \eqref{eq:dJstardissip1} -- is lost by dissipation within or outside the system. The net loss can thus be written as:
\begin{align}
\begin{split}
d\psi^* &= -dS^*|_{T,P,\{\mu_c\}} - \delta \sigma|_{T,P,\{\mu_c\}} 
\\& \qquad
- dS^*|_{S^*, \langle U \rangle, \langle V \rangle,\{\langle n_c \rangle\}} 
- \delta \sigma|_{S^*, \langle U \rangle, \langle V \rangle,\{\langle n_c \rangle\}} 
\\
&= -dS^* - \delta \sigma  \le 0
\end{split}
\label{eq:dJstardissip2}
\raisetag{13pt}
\end{align}
The terms $dS^*|_{S^*, \langle U \rangle, \langle V \rangle,\{\langle n_c \rangle\}}$ and $\delta \sigma|_{S^*, \langle U \rangle, \langle V \rangle,\{\langle n_c \rangle\}}$ constitute ``uncompensated transformations'' of Clausius \cite{Clausius_1865} -- respectively within and outside the system -- since they involve irreversible change(s) in one or more intensive variables.  For convenience, the four terms are unified in the second line of \eqref{eq:dJstardissip2}, into overall entropy changes within and outside the system.  

From \eqref{eq:dJstardissip2}, the system will approach the equilibrium position defined by $\psi^*=\psi^*_{min}$, equivalent to the maximum net irreversible increase in entropy within and outside the system. If $dS^* \ge 0$ and $\delta \sigma \ge 0$, this will again correspond to the state at which maximum thermodynamic entropy is produced, $\sigma=\sigma_{max}$. If the path of possible transitions is prescribed (e.g.\ constant $\{\langle f_r \rangle\}$, constant $\{\lambda_r\}$ or a mixed choice), the system will approach the local minimum $\psi^*_{min}=-\int (dS^*+\delta \sigma)$ along that path; if not, it will converge to the global minimum.

Thus in a dissipative, equilibrium system, the final equilibrium will occur at the position of minimum generalised potential $\phi^*_{eq}$ (maximum Massieu function $\lambda_0^*$).  If the incremental increases $dS^* \ge 0$ and $\delta \sigma \ge 0$ are unrecoverable, this will be equivalent to the position at which maximum thermodynamic entropy is produced. This provides an {\it equilibrium analogue} of the MEP principle, which is easily overlooked, since it involves a connection between equilibrium states and irreversible changes.

We can also consider the multiplier relations (\ref{eq:diff1}), variances (\ref{eq:diff2}) and covariances (\ref{eq:diff2mixed}), which yield a highly important set of relations for equilibrium systems, as listed in Table \ref{table_eq}. These include the ``Maxwell relations'' \cite{Maxwell_1888, Callen_1985}, which express the connections between various material properties or susceptibilities of the system (e.g.\ heat capacity, compressibility, coefficient of thermal expansion, activity coefficients, etc). The relations in Table \ref{table_eq} are valid only at equilibrium; i.e.\ they apply on the manifold of equilibrium positions.  Although discovered long before Jaynes' generic analysis, such relations demonstrate the power of Jaynes' method, particularly when applied to new situations.  

\begin{table}[t]
\caption{\label{table_eq} Multiplier relations (\ref{eq:diff1}), variances (\ref{eq:diff2}) and covariances (\ref{eq:diff2mixed}) for the example equilibrium thermodynamic system (\S\ref{equilsys}) at equilibrium (note $\var(a)=\langle a^2 \rangle - \langle a \rangle ^2$, $\cov(a,b)=\langle a b \rangle - \langle a \rangle \langle b \rangle$, and $\mu_b$ implies $b \ne c$). }
\begin{ruledtabular}
\begin{tabular}{llccc}
Multiplier relations  \\
$\langle U \rangle $ = $kT^2 \Bigl( \dfrac{\partial \ln \Xi^*}{\partial T} \Bigr)_{P,\{\mu_c\} }$ \vspace{5pt} \\
$\langle V \rangle $= $-kT \Bigl( \dfrac{\partial \ln \Xi^*}{\partial P} \Bigr)_{T, \{\mu_c\}}$ \vspace{5pt} \\
$\langle n_c \rangle $ = $kT \Bigl( \dfrac{\partial \ln \Xi^*}{\partial \mu_c} \Bigr)_{T, P, \{\mu_{b} \} }$ \vspace{5pt} \\
\hline 
{Variances}\\
$\var(U)$ = 
$kT^2  \Bigl( \dfrac{\partial \langle U \rangle}{\partial T} \Bigr)_{P, \{\mu_c\} }$  \vspace{5pt} \\
$\var(V)$ = 
$-kT  \Bigl( \dfrac{\partial \langle V \rangle}{\partial P} \Bigr)_{T, \{\mu_c\} }$  \vspace{5pt} \\
$\var(n_c)$ =
$kT  \Bigl( \dfrac{\partial \langle n_c \rangle}{\partial \mu_c} \Bigr)_{T, P, \{\mu_{b} \} }$  \vspace{5pt} \\
{Covariances (Maxwell relations)}\\
$\cov(U,V)$=
$- \Bigl( \dfrac{\partial \langle U \rangle}{\partial P} \Bigr)_{T, \{\mu_c\} }   = T  \Bigl( \dfrac{\partial \langle V \rangle}{\partial T} \Bigr)_{P, \{\mu_c\} }$  \vspace{5pt} \\
$\cov(U,n_c)$=
$\Bigl( \dfrac{\partial \langle U \rangle}{\partial \mu_c} \Bigr)_{T, P, \{\mu_b\} }   = T  \Bigl( \dfrac{\partial \langle n_c \rangle}{\partial T} \Bigr)_{P, \{\mu_c\} }$  \vspace{5pt} \\
$\cov(V,n_c)$=
$- \Bigl( \dfrac{\partial \langle n_c \rangle}{\partial P} \Bigr)_{T, \{\mu_c\} }   = \Bigl( \dfrac{\partial \langle V \rangle}{\partial \mu_c} \Bigr)_{T, P, \{\mu_b\} }$  \vspace{5pt} \\
$\cov(n_c,n_b)$ =
$\Bigl( \dfrac{\partial \langle n_c \rangle}{\partial \mu_b} \Bigr)_{T, P, \{\mu_{c \ne b} \}}   =   \Bigl( \dfrac{\partial \langle n_b \rangle}{\partial \mu_c} \Bigr)_{T, P, \{\mu_b\} }$  \vspace{5pt} \\
\end{tabular}
\end{ruledtabular}
\end{table}


\subsection{\label{Flow-Contr}Flow-Controlled (Steady-State) Systems}
We now examine a third kind of probabilistic system, which much more closely matches our experience of real systems and of life on Earth:\ the {\it flow-controlled system} (Figure \ref{fig:systems}c), constrained by flows of various physical quantities $f_r$ through the system. Flow-controlled systems are familiar to fluid mechanicists and engineers, as defined by the {\it control volume}, a geometric region through which fluid(s) can flow (the Eulerian description), bounded by its {\it control surface}. In the language of statistical mechanics, flow-controlled systems constitute a separate ``ensemble'', very different to those examined in equilibrium thermodynamics.  For the simplified flow system shown in Figure \ref{fig:systems}c, the mean rate of change of each conserved quantity $f_r$ with respect to time $t$ within the system is:
\begin{gather}
\frac {d  \langle f_r \rangle} {d t} = \langle {\flow{F}_r} \rangle_{in} - \langle {\flow{F}_r} \rangle_{out} + \langle {\dot{f}_r} \rangle_{prod}
\label{eq:flow_bal1} 
\end{gather}
where $\langle {\flow{F}_r} \rangle_{in}$, $\langle {\flow{F}_r} \rangle_{out}$ and $\langle {\dot{f}_r} \rangle_{prod}$ represent, respectively, the mean flow rates of $f_r$ into and out of the control volume, and rate of production within the control volume. At steady state, ${d  \langle f_r \rangle}/{d t} = 0$, and so for a system with no production terms, (\ref{eq:flow_bal1}) reduces to:
\begin{align}
\langle {\flow{F}_r} \rangle_{in} = \langle {\flow{F}_r} \rangle_{out} = \langle {\flow{F}_r} \rangle 
\label{eq:flow_bal2}
\end{align}
The mean flow rates $\langle {\flow{F}_r} \rangle$ can then be interpreted as constraints on the distribution of instantaneous flow rates $\flow{F}_{ri}$ through the system. By information-theoretic reasoning \cite{Shannon_1948, Shore_J_1980} and/or the combinatorial basis of entropy \cite{Boltzmann_1877, Planck_1901, Vincze_1972, Grendar_G_2001, Niven_CIT, Niven_MaxEnt07}, the MinXEnt principle can then be used to calculate the stationary or ``most probable'' distribution of flow rates within the system -- its {\it steady state position} -- subject to the constraints on the system. Moment constraints are just moment constraints, regardless of their physical manifestation.  In consequence, the entirety of Jaynes' generic approach (\S\ref{Jaynes}) can be applied to the analysis of steady state, as well as equilibrium, systems. Of course, flow-controlled systems may be subject to variable or cyclic flow rates, causing deviations from the steady state position; such effects are not considered further here.

Notwithstanding the broad applicability of Jaynes' method to both equilibrium and steady state systems, it is important to understand their differences. In equilibrium systems, the physical ``entities'' represent individual particles (e.g.\ molecules, ions, oscillators, etc), or entire systems of such particles, which can choose different values $f_{ri}$ of various physical parameters $f_r$. In contrast, in the Eulerian description of a steady state system, the ``entities'' represent individual fluid elements (in the limit, points), which may adopt particular instantaneous flow rates $\flow{F}_{ri}$ through the system. The latter analysis therefore examines the bulk behaviour of many individual particles passing through a control volume, which cannot readily be ascribed to particular particles. A further distinction is that equilibrium systems require no effort to maintain equilibrium, and can undergo reversible or irreversible changes; in contrast, maintenance or alteration of steady state systems inherently requires an irreversible change in the universe.  Aside from these differences, both systems have many features in common:
\begin{list}{$\bullet$}{\topsep 3pt \itemsep 3pt \parsep 0pt \leftmargin 8pt \rightmargin 0pt \listparindent 0pt
\itemindent 0pt}
\item In keeping with any method of inference, there is no guarantee that a predicted stationary position will occur, in that there may be (unknown) dynamic inhibitions (``non-ergodities'') which prevent its occurrence. Indeed, in equilibrium thermodynamics the existence of ``metastable'' states is well known, and is dealt with by kinetic theory (rate processes) and the activation energy concept. However, if the analysis incorporates everything that is known about the system, Jaynes' method will provide the ``best representation'' and ``expected position'' of the system \cite{Jaynes_1957}. In this vein, whilst it may be desirable to seek ``mechanistic'' or ``dynamic'' explanations for the observed steady state in particular flow-controlled systems -- analogous to justifications of equilibrium using the equations of motion \cite{Lanczos_1966} or Boltzmann's H-theorem \cite{Boltzmann_1872} -- the general principle must necessarily be inferential (statistical or probabilistic), just as it is in equilibrium thermodynamics.

\item For a given constraint set, the stationary position $D^*$ or $\mathfrak{H}^*$ 
is usually considered unique within its parameter (state function) space \cite{Jaynes_1957, Jaynes_1957b, Jaynes_1963, Tribus_1961a, Tribus_1961b, Kapur_K_1992, Jaynes_2003}.  However, this depends on the type of constraints; e.g.\ power-law constraints are known to give multiple stationary positions \cite{Grendar_G_2004a}. Uniqueness of the stationary position, in general, is therefore not claimed here, but is correct in the case of linear moment constraints \cite{Kapur_K_1992}.

\item A unique stationary position $D^*$ or $\mathfrak{H}^*$ in parameter space does not imply uniqueness of the dynamic structure(s) which can produce it; indeed, many such structures may be possible.  The occurrence of any given structure will depend on the interplay between the system history (hysteresis phenomena) and the dynamics (the extent of fluctuations, or ``jitter'', in the system), both of which lie outside the domain of Jaynes' method {\it per se}. This feature is also well known in equilibrium systems: e.g.\ the equilibrium position of a supersaturated solution does not provide any information about the particle size distribution or geometric form of the precipitant, which instead will depend on localised (history-dependent) nucleation, crystallisation and diffusion phenomena.

\item The MEP principle has been criticised for appearing to be a means of selection between a few isolated optima, rather than a true variational principle \cite{Bruers_2007c, Bruers_2007d}. However, many equilibrium systems also exhibit a discontinuous approach to equilibrium for dynamic reasons, especially those with a phase change; e.g.\ the rapid solidification of a saturated solution of sodium acetete trihydrate -- which can dissolve in its own water of crystallisation -- when tapped. The equilibrium between nitroglycerine and oxygen also illustrates this point. In flow-controlled systems, we might expect any locally optimal steady states $\{ \phi^*_{opt} \}$ to be similarly narrowly defined, since they may require special coordination of flows in different domains. Thus although derived by variational means, there is nothing wrong with the minimisation of $\phi^*$ as a selection principle.  

\end{list}
\noindent The rationale for the analysis of a steady state system by Jaynes' method is therefore identical -- with many of the same caveats -- to that for equilibrium systems.
%

\section{\label{steady_state}Analysis of Flow-Controlled Systems} 

\subsection{\label{st-CV}Control Volume Analysis}
%
\begin{figure}[t]
\begin{center}
\setlength{\unitlength}{0.6pt}
  \begin{picture}(380,320)
   \put(0,0){\includegraphics[width=75mm]{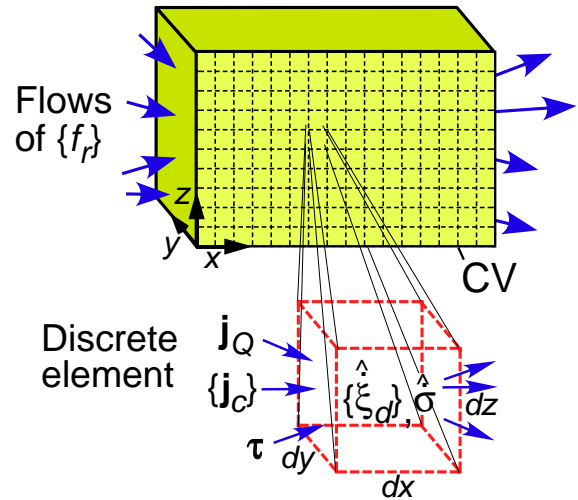} }
  \end{picture}
\end{center}
\caption{{(Color online)} Discretised control volume model of a flow-controlled thermodynamic system.}
\label{fig:CV}
\end{figure}

Calculation of the steady state position of a flow-controlled system can now proceed, firstly involving a traditional engineering thermodynamic analysis.  Consider the control volume shown in Figure \ref{fig:CV}, divided into infinitesimal volume elements, each of which is assumed to satisfy the ``local equilibrium assumption''.  This allows the definition of local specific (per unit mass) quantities including specific internal energy $u$, specific volume $\rho^{-1}$, specific moles $m_c$ of each species $c$ and specific entropy $s$ of each element, where $\rho$ is the fluid density, as well as intensive variables such as temperature $T$, pressure $P$ and molar chemical potentials $\mu_c$ of each species $c$ on its boundary. The elements are thus sufficiently small for local equilibrium to hold, but obviously represent averages over regions large enough to be considered a thermodynamic continuum. In general, for an element subject to the total entropy flux\footnote{We adhere {\it strictly} to engineering convention: a flow is measured in SI units of quantities s$^{-1}$, whilst a flux is expressed in quantities m$^{-2}$ s$^{-1}$.} $\vect{j}_{s,tot}$, the volumetric rate of entropy production within the element (SI units:\ J K$^{-1}$ m$^{-3}$ s$^{-1}$) and overall rate of production (J K$^{-1}$ s$^{-1}$) are \cite{deGroot_M_1962, Prigogine_1967, Kreuzer_1981, Bird_etal_2006}:
\begin{align}
\hat{\dot{\sigma}} &=\frac{\partial \rho s}{\partial t} +  {\vect \nabla} \cdot {\vect j}_{s,tot} \ge 0
\label{eq:sigma_dot_hat}
\\
{\dot{\sigma}} &=\iiint_{CV} \hat{\dot{\sigma}} \, dV \ge 0
\label{eq:sigma_dot_tot}
\end{align}
where $\vect{\nabla}=[\partial / \partial x, \partial / \partial y, \partial / \partial z]^\top$ is the Cartesian gradient operator; $\vect{a} \cdot \vect{b}$ is the vector scalar product (dot product); and $\vect{a}^\top$ is the transpose of $\vect{a}$. For an element with mean local heat flux ${\vect j}_Q$, diffusive mass fluxes ${\vect j}_c$ of each species $c$ (relative to the mass-average velocity ${\vect v}$ through the element), viscous stress tensor\footnote{The viscous stress tensor ${\tens {\tau}}$ is equal to the net or molecular stress tensor ${\tens {\Pi}}$ less the uniform pressure component $P$, whence $\tens{\tau}=\tens{\Pi}-P\tens{\delta}$, where ${\tens {\delta}}$ is the Kronecker delta tensor \cite{deGroot_M_1962, Kreuzer_1981, Bird_etal_2006}. The stress notation of Bird {\it et al.} \cite{Bird_etal_2006} is adopted here, in which $P>0$ and $\tau_{ij}>0$ denote compression.} ${\tens{\tau}}$ and molar rate per unit volume $\hat{\dot{\xi}}_{d}$ of each chemical reaction $d=1,...,D$, from the conservation of energy, mass,  momentum and chemical species, the total entropy flux and rate of entropy production per volume reduce to \cite{deGroot_M_1962, Prigogine_1967, Kreuzer_1981, Bird_etal_2006}:
\begin{gather}
{\vect{j}}_{s,tot} = \rho s {\vect v} + {\vect j}_Q \biggl( \frac {1}{T} \biggr) - \sum\limits_{c=1}^C {\vect j}_c \biggl( \frac{{\mu}_c}{M_c T} \biggr) 
\label{eq:s_flux}
\\
\begin{split}
\hat{\dot{\sigma}}= 
{\vect j}_Q \cdot {\vect \nabla} \biggl( {\frac{1}{T}} \biggr) 
- \sum\limits_{c=1}^C {{\vect j}_c} \cdot \biggl[ {\vect \nabla} \biggl( \frac{\mu_c}{M_c T} \biggr) - \frac {{\vect g}_c}{T}   \biggr]
\\
 -  {{\tens{\tau}}} : { \vect \nabla} \biggl( \frac{ { \vect v} }{T} \biggr)^\top
 -  \sum\limits_{d=1}^{D} \hat{\dot{\xi}}_{d} \frac{A_{d}}{T}  \quad \ge 0
\end{split}
\label{eq:sigma_dot_hat2}
\end{gather}
where $\tens{A} : \tens{B}$ is the tensor scalar product; $\vect{a} \vect{b}^\top$ is the vector dyadic product (often simply written $\vect{a} \vect{b}$); $M_c$ is the molecular mass of species $c$; ${\vect g}_c$ is the specific body force on each species $c$; and $A_{d}=\sum\nolimits_{c=1}^{C} \nu_{cd} \, {\mu}_c$ is the molar chemical affinity of reaction $d$, wherein $\nu_{cd}$ is the stoichiometric coefficient of species $c$ in the $d$th reaction ($\nu_{cd}>0$ for a product and $A_{d}<0$ denotes spontaneous forwards reaction). Alternative formulations of \eqref{eq:sigma_dot_hat2} are available for different situations \cite{deGroot_M_1962, Kreuzer_1981, Bird_etal_2006}; the present formulation is sufficiently broad for the present discussion.

In engineering modelling, the local mass, momentum and energy conservation equations and local entropy production \eqref{eq:sigma_dot_hat2} are usually non-analytic, and so are calculated on a {two- or three-dimensional grid} of finite rather than infinitesimal domains (e.g., by the finite element method) such that the local solutions are self-consistent and match the boundary conditions and source-sink terms of the overall control volume \cite{deGroot_M_1962, Kreuzer_1981, Bird_etal_2006, Moran_S_2006, White_2005}. This approach assumes that the calculated solution is unique. However, for many flow systems, the entropy production \eqref{eq:sigma_dot_hat2} is indeterminate, since one or more of the fluxes (${\vect j}_Q$, ${\vect j}_c$, ${{\tens{\tau}}}$ and $\hat{\dot{\xi}}_{d}$) and/or gradients (${\vect \nabla} (T^{-1})$, ${\vect \nabla} ({\mu}_c / T)$, ${\vect{g}}_c/T$, ${\vect \nabla} ({\vect v/T})^\top$ and $A_{d}/T$) across the overall control volume are unknown, there being no principle within equilibrium thermodynamics by which they can be determined. 
In such cases, \eqref{eq:sigma_dot_hat2} has been solved by assuming the near-equilibrium linear (Onsager) regime, for which closure of the equations can be attained by the use of the equilibrium Gibbs-Duhem relation, Curie postulate, tensor symmetry and specification of the (linear) mass, momentum and energy diffusion coefficients \cite{deGroot_M_1962, Kreuzer_1981, Bird_etal_2006}. Outside the linear regime, such as in turbulent fluid flow, these assumptions can produce serious computational difficulties (e.g.\ the need for extremely small elements) and are commonly handled by broad simplifications or empirical rules. {Indeed, for many flow systems, the entropy production is completely disregarded.}  
Eqs.\ \eqref{eq:sigma_dot_hat}-\eqref{eq:sigma_dot_hat2} therefore apply to a flow-controlled thermodynamic system at steady state, but in many systems, may not uniquely define the steady state. 

\subsection{\label{st-MinXEnt}Jaynes Analysis}
We now examine a flow-controlled system from Jaynes' perspective (\S\ref{Jaynes}).  In general, for an infinitesimal fluid element of a control volume, Jaynes' relations for the steady state position
are exactly as given in (\ref{eq:pstar2_i})-(\ref{eq:Massieu}), but this time with mean flux constraints $ { {\vect j}_r } $ instead of quantity constraints $\langle  { {f}_r }  \rangle$ (for convenience the $\langle \cdot \rangle$ notation is dropped), and increments in generalised heat fluxes {$\delta \vect{q}_r$} instead of in the generalised heats $\delta Q_r$. The Lagrangian multipliers, say $\zeta_r$, will of course be different to those of equilibrium thermodynamics $\lambda_r$, conjugate to the flux constraints considered. We should also denote the cross-entropy or entropy by some special symbol, e.g.\ $D_{st}$ or $\mathfrak{H}_{st}$, to emphasise that it applies to a steady state system, and is quite different to that for an equilibrium system, $D_{eq}$ or $\mathfrak{H}_{eq}$. Our generic analysis of the steady state position of a flow-controlled system is now complete, requiring only the substitution of relevant parameters for the system under consideration.

Returning to the example of \S\ref{st-CV}, we consider that each element can experience instantaneous values of the heat flux ${\vect j}_{Q,\flow{I}}$, mass fluxes ${\vect j}_{\flow{N}_c}$ of each species $c$, stress tensor ${{\tens{\tau}}}_{\flow{J}}$ and reaction rates $\hat{\dot{\xi}}_{\flow{L}_{d}}$, where the indices $\flow{I},\flow{J}, \flow{L}_{d}$ and $\flow{N}_c$ can (in principle) take any value. The system is constrained by mean values of the heat flux ${\vect j}_Q$, mass fluxes ${\vect j}_{c}$, stress tensor ${\tens{\tau}}$ and reaction rates $\hat{\dot{\xi}}_d$ through or within the element, and by the natural constraint (\ref{eq:C0}), where expectations are taken with respect to the joint probability $\pi_{\vecti} =  \pi_{\flow{I},\flow{J},\{\flow{L}_d\},\{\flow{N}_c\} }$ of instantaneous fluxes.  As for the equilibrium analysis (\S\ref{equilsys}) we assume equiprobable states, and thus maximise the Shannon entropy:
\begin{gather}
\begin{split}
\mathfrak{H}_{st} =
-  \sum\limits_{\substack{ \flow{I}=\\ \;\;-\infty}}^{\infty}   \sum\limits_{\substack{ \flow{J}=\\ \;\;-\infty}}^{\infty}     \sum\limits_{\substack{ \flow{L}_{1}=\\ \;\;-\infty}}^\infty   ... \sum\limits_{\substack{ \flow{L}_{D}=\\ \;\;-\infty}}^\infty  \sum\limits_{\substack{ \flow{N}_{1}=\\ \;\; -\infty}}^\infty   ... \sum\limits_{\substack{ \flow{N}_C=\\ \;\;-\infty}}^\infty
\pi_{\vecti} \ln \pi_{\vecti} ,
\end{split}
\label{eq:H_st}
\end{gather}
subject to its constraints. From (\ref{eq:pstar2_i})-(\ref{eq:MinXEnt}) and (\ref{eq:dDstar}):
\begin{gather}
\begin{split}
\pi_{\vecti}^* &= \dfrac {\beta_{\vecti}^*  }  {\flow{Z}^*}, 
\\
\beta_{\vecti}^* &= \exp{  \biggl(  - \vect{\zeta}_{Q} \cdot {\vect j}_{Q,\flow{I}} - \sum\limits_{c=0}^{C} \vect{\zeta}_c \cdot {\vect j}_{ \flow{N}_c} } - {\tens \zeta}_{\tens \tau} : {\tens \tau}_{\flow{J}} 
 - \sum\limits_{d=1}^{D} \zeta_{d} \hat{\dot{\xi}}_{\flow{L}_d} \biggr)
\\
\flow{Z}^* &= e^{\zeta_0^*} =
\sum\limits_{\substack{ \flow{I}=\\ \;\;-\infty}}^{\infty}   \sum\limits_{\substack{ \flow{J}=\\ \;\;-\infty}}^{\infty}     \sum\limits_{\substack{ \flow{L}_{1}=\\ \;\;-\infty}}^\infty   ... \sum\limits_{\substack{ \flow{L}_{D}=\\ \;\;-\infty}}^\infty  \sum\limits_{\substack{ \flow{N}_{1}=\\ \;\; -\infty}}^\infty   ... \sum\limits_{\substack{ \flow{N}_C=\\ \;\;-\infty}}^\infty
\beta_{\vecti}^*
\end{split}
\label{eq:pi_star}
\raisetag{28pt}
\\
 \mathfrak{H}_{st}^*  = \ln \flow{Z}^*  + \vect{\zeta}_Q \cdot  {\vect j}_Q + \sum\limits_{c=1}^C \vect{\zeta}_c \cdot {\vect j}_{c} + \tens{\zeta}_{\tens{\tau}} : {\tens{\tau}} + \sum\limits_{d=1}^D \zeta_d \hat{\dot{\xi}}_d
 \label{eq:MaxEnt_st}
\\
{d\mathfrak{H}_{st}^*   = \vect{\zeta}_Q \cdot  \delta \vect{q}_Q + \sum\limits_{c=1}^C \vect{\zeta}_c \cdot \delta \vect{q}_c + \tens{\zeta}_{\tens{\tau}} : \delta \tens{q}_{\tens{\tau}} + \sum\limits_{d=1}^D \zeta_d \delta q_d }
\label{eq:dHstar_st}
\end{gather}
where the multipliers ${\vect \zeta}_{Q}$, ${\vect{\zeta}}_c$, ${\tens{\zeta}}_{\tens {\tau}}$ and $\zeta_d$ are associated with the heat, species, stress tensor and reaction constraints (${\vect \zeta}_{Q}$ and ${\vect{\zeta}}_c$ are vectors and ${\tens{\zeta}}_{\tens {\tau}}$ a second order tensor); {$\delta \vect{q}_Q$, $\delta \vect{q}_c$, $\delta \tens{q}_{\tens{\tau}}$ and $\delta q_d$} are the corresponding changes in generalised heat fluxes; $\zeta_0^*$ is the Massieu function and $\flow{Z}^*=e^{\zeta_0^*}$ the partition function. We see that $\mathfrak{H}_{st}^*$ is a dimensionless flux entropy. {Eqs.\ \eqref{eq:Massieu} and \eqref{eq:MaxEnt_st}-\eqref{eq:dHstar_st} then give the change in a generalised potential function for steady state systems:}
\begin{align}
\begin{split}
&d \phi^*_{st} = - d \zeta_0^*  = - d \ln \flow{Z}^* \\
&=  - d \mathfrak{H}_{st}^*  + \vect{\zeta}_Q \cdot  d {\vect j}_Q + \sum\limits_{c=1}^C \vect{\zeta}_c \cdot d {\vect j}_{c} + \tens{\zeta}_{\tens{\tau}} : d {\tens{\tau}} + \sum\limits_{d=1}^D \zeta_d d \hat{\dot{\xi}}_d \\
& \quad + d \vect{\zeta}_Q \cdot  {\vect j}_Q + \sum\limits_{c=1}^C d \vect{\zeta}_c \cdot  {\vect j}_{c} + d\tens{\zeta}_{\tens{\tau}} : {\tens{\tau}} + \sum\limits_{d=1}^D d\zeta_d \hat{\dot{\xi}}_d
\end{split}
\label{eq:Massieu_st}
\raisetag{21pt}
\end{align}
Again -- with Gibbs -- we could exclude a net change in the multipliers $\zeta_r$, giving $d \phi^*_{st}|_{\{\zeta_r\}} = \sum_{r=1}^R \zeta_r \delta \vect{w}_r$, where $\delta \vect{w}_r$ are changes in generalised work fluxes, in addition to those {incorporated by other} constraints. However, as with equilibrium systems (\S\ref{equilsys}), we wish to consider all possible variations of $\phi^*_{st}$.  Integration of \eqref{eq:Massieu_st}, from the zero flux, zero multiplier (equilibrium) position $\phi^*_{st0} = - \mathfrak{H}^*_{st0}$ yields: 
\begin{align}
\begin{split}
&\phi_{st}^*  = - \mathfrak{H}_{st}^*    + \vect{\zeta}_Q \cdot  {\vect j}_Q + \sum\limits_{c=1}^C \vect{\zeta}_c \cdot {\vect j}_{c} + \tens{\zeta}_{\tens{\tau}} : {\tens{\tau}} + \sum\limits_{d=1}^D \zeta_d \hat{\dot{\xi}}_d
\end{split}
\label{eq:Massieu_st2}
\end{align}
This is the steady state analogue of the equilibrium generalised potential $\phi^*_{eq}$. Relations \eqref{eq:pi_star}-\eqref{eq:Massieu_st2} apply only to quasistatic transitions ``on the manifold of steady state positions''. However, in this case -- since this is a flow system -- they always involve irreversible processes. 

\subsection{\label{Synthesis}Synthesis}

We can now combine the control volume (\S\ref{st-CV}) and Jaynesian (\S\ref{st-MinXEnt}) analyses of the example problem (Figure \ref{fig:CV}).  Since the fluxes in \eqref{eq:sigma_dot_hat2} and  \eqref{eq:Massieu_st2} are linearly independent (not necessarily orthogonal), it is possible to equate similar terms, giving the following identities:
\begin{align}
\phi^*_{st} &= - \mathfrak{H}_{st}^* - \frac {\tscale \Vscale}{k} \hat{\dot{\sigma}}  
\label{eq:ass_phi_st}  
\\
 {\vect{\zeta}}_Q &= - \frac {\tscale \Vscale}{k} {\vect \nabla} \biggl( {\frac{1}{T}} \biggr) 
  \label{eq:ass_zeta_Q}
\\
 {\vect{\zeta}}_c &= \frac {\tscale \Vscale}{k} \biggl[ {\vect \nabla} \biggl( \frac{\mu_c}{M_c T} \biggr) - \frac {{\vect g}_c}{T}   \biggr]
  \label{eq:ass_zeta_c}
\\
 {\tens{\zeta}}_{\tens{\tau}} &=  \frac {\tscale \Vscale}{k} { \vect \nabla} \biggl( \frac { \vect v }{T} \biggr)^\top
\label{eq:ass_zeta_Pi}
\\
 \zeta_d &= \frac {\tscale \Vscale}{k} \frac {A_d}{T}
 \label{eq:ass_zeta_d}
\end{align}
where $\tscale$ and $\Vscale$ respectively are characteristic time and volume scales for the system considered, which emerge from the fact that $\phi^*_{st}$, $\mathfrak{H}_{st}^*$ and each product in \eqref{eq:Massieu_st2} must be dimensionless\footnote{An alternative view, not considered further here, is to interpret $(\tscale \Vscale /k) \, (\vect{g}_c / T) \cdot \vect{j}_c $ as a generalised work flux, which must be subtracted from a redefined generalised heat flux $\delta \vect{q}_{\vect{j}_c}' = (\tscale \Vscale /k) \, \vect{\nabla} (\mu_c / M_c T) \cdot \vect{j}_c$.}$^{,}$\footnote{Note $\tscale \Vscale$ has SI units of m$^3$ s, equivalent to a four-dimensional space-time element, or action divided by pressure.}.  

From \eqref{eq:ass_phi_st} and the foregoing analysis (\S\ref{Jaynes} and \S\ref{equilsys}), if the flow-controlled system contains some process(es) which dissipate its generalised potential $\phi^*_{st}$, it will move by increments:
\begin{align}
\begin{split}
d \phi^*_{st} &= - d \mathfrak{H}_{st}^* - \frac {\tscale \Vscale}{k} \delta \hat{\dot{\sigma}}   \le 0
\end{split} 
\label{eq:diff_ass_phi_st}
\end{align}
until it converges to a final steady state position given by $d \phi^*_{st}=0$, hence at minimum $\phi^*_{st}$. Eq.\ \eqref{eq:diff_ass_phi_st} is analogous to \eqref{eq:dJstardissip2} for equilibrium systems, accounting for the loss of generalised potential within and outside the system. If $d \mathfrak{H}_{st}^* \ge 0$ and $\delta \hat{\dot{\sigma}} \ge 0$, the steady state position $\phi^*_{st}=\phi^*_{st,min}$ will correspond to the position of maximum entropy production $\hat{\dot{\sigma}}=\hat{\dot{\sigma}}_{\max}$.  The analysis therefore provides a theoretical justification for a local form of the MEP principle, conditional on the assumption that incremental increases in $ \mathfrak{H}_{st}^*$ and $\hat{\dot{\sigma}}$ are unrecoverable.  As with equilibrium systems (\S\ref{equilsys}), the dissipation may be confined to a constant-multiplier $\{\zeta_r\}$,  constant-constraint $\{\vect{j}_r\}$ or mixed constraint path, leading to a local (path-dependent) minimum $\phi^*_{st,min}$, or may be free to choose its own path, whereupon it will attain a global minimum.
 
Processes for which $\delta \hat{\dot{\sigma}}>0$ (to which the MEP principle applies) are here termed {\it exoentropogenic}\footnote{From ancient Greek:\ {\it exo-}, outer or external (obverse {\it endo-}, within or internal); {\it tropos}, transformation (used by Clausius \cite{Clausius_1865}); and {\it -genic}, generating or producing.}, since they result in the production of entropy and its export to the rest of the universe. Exoentropogenic processes are one class of processes leading to the formation of steady-state flow systems, in the same way that {\it exothermic} reactions -- in which the heat $\delta Q<0$ -- are one class of processes which generate equilibrium systems.  It is left as an open question here whether exoentropogenic processes (and hence the MEP principle) are universal in application, or are merely an important class of processes applicable to flow systems at steady state.
  
\subsection{\label{Imps}Implications}

The analysis has several important implications. Firstly, under conditions in which the MEP principle applies, from \eqref{eq:sigma_dot_hat2} and \eqref{eq:ass_phi_st}:
\begin{list}{$\bullet$}{\topsep 0pt \itemsep 3pt \parsep 3pt \leftmargin 8pt \rightmargin 0pt \listparindent 0pt
\itemindent 0pt}
\item If the fluxes or rates ${\vect j}_Q$, ${\vect j}_c$, ${{\tens{\tau}}}$ and $\hat{\dot{\xi}}_{d}$ in \eqref{eq:sigma_dot_hat2} are indeed constant, the system will achieve MEP by maximising the magnitudes of the gradients or forces $|{\vect \nabla} (T^{-1})|$, $|{\vect \nabla} ({\mu}_c / T)|$, $|{\vect{g}}_c/T|$, $|{\vect \nabla} ({\vect v/T})^\top|$ and $|A_{d}/T|$ across or within each element, each weighted by its conjugate flux term. Since the gradients can be interpreted as measures of {disequilibrium} of the system, we see that a flow-controlled system is driven to a steady state position of {\it maximum disequilibrium}. 
\item If, on the other hand, the gradients or forces ${\vect \nabla} (T^{-1})$, ${\vect \nabla} ({\mu}_c / T)$, ${\vect{g}}_c/T$, ${\vect \nabla} ({\vect v/T})^\top$ and $A_{d}/T$ in \eqref{eq:sigma_dot_hat2} are fixed, the system will be driven to maximise the fluxes or rates ${\vect j}_Q$, ${\vect j}_c$, ${{\tens{\tau}}}$ and $\hat{\dot{\xi}}_{d}$, again in the form of a weighted sum.  In other words, the most probable response of a system to gradients or forces in one or more intensive variables $\lambda_r$ is {\it the occurrence of flow} of the corresponding physical quantities $f_r$, at their maximum mean flow rates. This is also consistent with maximum disequilibrium of the system. 
\end{list}
The distinction between fixed flux ${\vect j}_r$ (Neumann) or intensive variable gradient ${\vect \nabla} \lambda_r$ (Dirichlet) boundary conditions is well known in the solution of differential equations. In the MaxEnt analysis of a steady state system, the choice between such boundary conditions plays exactly the same role as the choice between fixed content $\langle f_r \rangle$ or intensive variable $\lambda_r$ constraints in an equilibrium system, or in other words, between corresponding microcanonical and canonical representations. Although the two representations within each pair are different, the mathematical apparatus used to examine either set of constraints is the same \citep[c.f.][]{Jaynes_1957}. Flow-controlled systems can also be subject to composite flux-gradient (Robin) boundary conditions, or a mixed set of conditions, giving rise to steady states which have no equilibrium analogue in quantity-constrained systems.

To achieve maximum $\hat{\dot{\sigma}}$ in \eqref{eq:sigma_dot_hat2}, the conjugate vectorial fluxes and gradients should be collinear.  A consequence of the MEP principle is that the free fluxes or gradients will try to orient themselves to attain collinearity, to the maximum extent allowable by competition between different flows and/or by anisotropy within the system (e.g.\ due to the velocity gradient). Similarly, the system will endeavour to align its principal stresses with the principal directions of the velocity gradient. This may provide an explanation for the many simplifications used in control volume analysis, such as tensor symmetry.

Whilst \eqref{eq:sigma_dot_hat2} is in differential form, in systems with simple boundary conditions it may be possible to apply its finite difference approximation to much larger domains. This appears to be the case for the Earth climate system, which has been analysed (on a whole-Earth scale) using simple two-box or ten-box models \cite{Paltridge_1975, Paltridge_1978, Paltridge_1981, Ozawa_Ohmura_1997, Paltridge_2001, Shimokawa_O_2001, Shimokawa_O_2002, Kleidon_etal_2003, Ozawa_etal_2003, Kleidon_L_book_2005, Kleidon_L_art_2005, Davis_2008, Paltridge_etal_2007}. {The analysis also indicates MEP to be a local principle, which need not apply to the universe as a whole. Only the entropy production of a system -- not the entire universe -- is maximised, since it is only within a system that there exist mechanisms by which the MEP state can be attained.  This is a characteristic feature of previous applications of the MEP principle}, which has incited much debate \cite{Paltridge_1975, Paltridge_1978, Paltridge_1981, Ozawa_Ohmura_1997, Ozawa_etal_2003, Kleidon_L_book_2005, Bruers_2007c}. 

Tidying up the analysis, \eqref{eq:ass_phi_st}-\eqref{eq:ass_zeta_d} can be substituted into \eqref{eq:pi_star}-\eqref{eq:Massieu_st2}.  Steady state analogues of the four laws of equilibrium thermodynamics can also be obtained from Jaynes' approach \cite{Tribus_1961b, Kapur_K_1992}, as set out in Appendix \ref{Apx_laws}. We can also consider the set of multiplier relations (\ref{eq:diff1}), variances (\ref{eq:diff2}) and covariances (\ref{eq:diff2mixed}) for steady state systems, examined further in Appendix \ref{Apx_rels}. The analysis therefore provides a set of {\it testable} laws and transport relations for the behaviour of steady state thermodynamic systems. {Furthermore, as shown in Appendix \ref{Apx_rels}, the Onsager linear regime emerges as a first-order approximation to the analysis, in the vicinity of equilibrium. The analysis therefore provides a general mathematical formalism for the analysis of flow systems -- including non-equilibrium thermodynamic systems -- both near and far from equilibrium.}
 

This is as far as our comparative analysis of steady state thermodynamics can take us, using the parameters of equilibrium thermodynamics, and it is important that it not be taken too far. For example, in many systems the local equilibrium assumption -- which forces the MEP principle to adopt a local formulation -- may be too restrictive; such considerations lead into the domain of more complicated approaches, such as extended irreversible thermodynamics \cite{Jou_etal_1993, Jou_etal_1999}.  For more comprehensive treatments it may be necessary to abandon the connection with equilibrium parameters, and conduct the steady state analysis using the ``raw'' Lagrangian multipliers $\zeta_r$ -- or some functions thereof -- obtained directly by MinXEnt.  The raw multipliers (the ``ideal'' gradients) could then be correlated to the actual gradients by {\it gradient coefficients}, analogous to the activity and fugacity coefficients of equilibrium thermodynamics. In some systems, it may even be necessary to adopt process- and direction-specific time and volume scales $\tscale_{r\imath\jmath}$ and $\Vscale_{r\imath\jmath}$ (possibly also varying with position), leading to weighted scales $\tscale_{sys}$ and $\Vscale_{sys}$ for the system as a whole. The analysis extends naturally to further developments, for example the use of a ``local steady state assumption'' in the analysis of transient phenomena. This leads into a higher-order theory of acceleration phenomena, in which a maximum is sought in the sum of products of accelerative transport terms $\partial {\vect j}_r / \partial t$ and gradients of the gradients $\vect{\nabla} \vect{\zeta}_r^\top$ (a conditional, maximum $\hat{\ddot{\sigma}}$ principle) \cite{Crucifix_2007_pc}, an idea best examined elsewhere.

\section{\label{Discussion}Discussion}
The foregoing analysis would be incomplete without a brief account of its implications. From \S\ref{steady_state}, MEP emerges as a {local, conditional} principle, in which each local control volume behaves as an {\it actor} or {\it agent} which seeks to minimise its generalised potential $\phi^*_{st}$, and hence (conditionally) to maximise its local entropy production.  This provides a  driving force for the formation and reinforcement of ``emergent'' self-organisation of the system as a whole, since by such ``cooperative federalism'', each local ``selfish'' control volume can achieve much higher entropy production than it could of its own volition.  We therefore see that the MEP principle drives the formation of {\it structure} and {\it function}.  Of course, this does not in any way preclude the development of {\it competition} between control volumes -- or even between associations of control volumes or entire systems -- for a greater share of generalised potential; such competitive forces are certainly well known to us. These twin effects, a predominant, higher level cooperation and a lower level (but occasionally overwhelming) competition, are the hallmark of the ``dynamic steady state'' of a dissipative, flow-controlled system.  

The analysis therefore confirms a number of aspects of the philosophy of Prigogine and co-workers on non-equilibrium, dissipative systems \cite{Prigogine_1967, Prigogine_1980, Prigogine_S_1984, Kondepudi_P_1998}, even though his ``minimum entropy production principle'' (valid only in the linear regime) is quite different \cite{Ozawa_etal_2003, Martyushev_S_2006}.  It also provides an explanation for the  ``constructal law'' of Bejan \cite{Bejan_1997d}: \noindent {\it ``For a finite size flow system to persist in time (to survive) its configuration must evolve .. [to] provide easier and easier access to the currents that flow through it.''} Furthermore, it confirms the essence of the (non-mathematical) gradient theory of Schneider and Sagan \cite{Schneider_S_2005}, who argue that {{\it ``Nature abhors a gradient''}; i.e. the occurrence of flow is a natural response to a physical gradient}.

Finally, the analysis {goes to the heart of} the ``riddle of life'' posed by many scientists \cite[e.g.][]{Schrodinger_1944, Kauffmann_2004}, concerning the perceived contradiction between the second law of thermodynamics and the existence of life. To recap:\ it is one thing to suggest that life {\it can} form, in that its ability to increase the thermodynamic entropy of the universe exceeds the reduction of entropy associated with its structure. But if life were merely an accident -- a fluctuation -- why should it not just die out? Why should it be so resilient to extreme events, as evidenced by its regrowth after the many mass-extinction episodes in the history of the Earth? As shown, the existence and evolution of life is driven by a deeper, purely probabilistic form of the second law \cite{Niven_MaxEnt07}: \\

``{\it A system tends towards its most probable form.}'' \\

\noindent This ``MaxProb'' principle \cite{Boltzmann_1877, Planck_1901, Vincze_1972, Grendar_G_2001, Niven_CIT, Niven_MaxEnt07}, applied to an isolated or open, dissipative equilibrium system, drives the system towards the equilibrium position $\mathfrak{H}^*_{eq}$ or $S^*$ of minimum generalised potential $\phi^*_{eq}$ (e.g.\ minimum free energy $F^*, G^*$ or $J^*$), since this is more probable than other forms of the system.  In contrast, in a flow system it {provides a driving force for} the formation of complex, dissipative structures, including life, to attain the local steady state position $\mathfrak{H}^*_{st}$ of minimum generalised potential $\phi^*_{st}$, hence (conditionally) of maximum $\hat{\dot{\sigma}}$, since this position -- not the equilibrium position -- is more probable than other realisations of the system. The above probabilistic statement of the second law therefore contains within it both a ``force of death'' and a ``force of life'', associated respectively with constraints on the mean contents $\langle f_r \rangle$ or fluxes $\langle \vect{j}_r \rangle$ (or their corresponding multipliers). It also implies a surprising inevitability to the evolution of life in a flow-controlled system, whenever the conditions are suitable, and indeed, of other complex systems such as fluid turbulence, biological and ecological structures, transport and communications networks, economic systems and human (or sentient) civilisation. 

\section{\label{Concl}Conclusions}
In this study, a clear distinction is made between (i) {\it quantity-controlled systems}, constrained by a set of mean physical quantities $\langle f_r \rangle$ and/or their corresponding multipliers $\lambda_r$, which converge towards an {\it equilibrium position}, and (ii) {\it flow-controlled systems}, as defined by a control volume (the Eulerian description), constrained by a set of mean fluxes $\langle \vect{j}_r \rangle$ and/or their corresponding multipliers $\zeta_r$, which converge towards a {\it steady state position}. A theory to determine the steady state position of a flow-controlled thermodynamic system is derived using the generic MaxEnt principle of Jaynes \cite{Jaynes_1957, Jaynes_1957b, Jaynes_1963, Tribus_1961a, Tribus_1961b, Shore_J_1980, Kapur_K_1992, Jaynes_2003}.  The analysis is shown to yield a local, conditional form of the MEP principle. It also yields steady state analogues of the four laws of equilibrium thermodynamics, and sets of multiplier, variance and covariance (Maxwell-like) relations applicable to flow-controlled systems at steady state.  The derivation is limited to reproducible flow-controlled systems, but does not appear to be restricted to the near-equilibrium linear regime; indeed, the latter can be recovered as a special case {(see Appendix \ref{Apx_rels})}.  The analysis reveals a very different manifestation of the second law of thermodynamics in steady state systems, which {provides} a driving force for the formation of complex, dissipative systems, including life.

Further work is required on the scope of the present analysis, and the relationship between concepts developed here and those of others; e.g.\ between the steady state generic entropy $\mathfrak{H}^*_{st}$, the path-based entropies of Dewar \cite{Dewar_2003, Dewar_2005} and Attard \cite{Attard_2006a, Attard_2006b} {and the quantum formulation of Beretta \cite{Beretta_2001, Gyft_Beretta_2005, Beretta_2006, Beretta_2008}}. Greater attention should also be paid to the tremendous power of the MinXEnt / MaxEnt method pioneered by Jaynes \cite{Jaynes_1957, Jaynes_1957b, Jaynes_1963, Tribus_1961a, Tribus_1961b, Kapur_K_1992, Jaynes_2003}, and its combinatorial (probabilistic) basis \cite{Boltzmann_1877, Planck_1901, Vincze_1972, Grendar_G_2001, Niven_CIT, Niven_MaxEnt07}.

\section*{Acknowledgments}
The author warmly thanks the participants of the 2007 and 2008 MEP workshops hosted by the Max-Planck-Institut f\"ur Biogeochemie, Jena, Germany, especially Axel Kleidon, Roderick Dewar, Filip Meysman, Graham Farquhar, Michel Crucifix, Peter Cox, Tim Jupp, Adrian Bejan and Ralph Lorenz, for valuable discussions; Bjarne Andresen, Marian Grendar, Bernd Noack, many of the above {and the anonymous reviewers} for comments on the manuscript; The University of New South Wales and the above Institute for financial support; and the European Commission for financial support as a Marie Curie Incoming International Fellow (contract 039729-rkniven-mc-iif-06) under Framework Programme 6.

\appendix

\section{\label{Apx_termin} Terminology}  

From the foregoing discussion (especially \S\ref{Jaynes}-\S\ref{steady_state}), it is evident that there are difficulties in terminology concerning the negative Massieu function $\phi^*=-\lambda_0^*=-\ln Z^*_q$.  Although described herein as a free energy-like concept or ``generalised potential'', from \eqref{eq:MinXEnt} it has more in keeping with an entropy-related quantity.  In equilibrium systems, $\phi^*_{eq}$ is usually multiplied by $kT$ to give the free energy; but as shown in \S\ref{equilsys}, it is more appropriate to multiply it by $k$, to avoid complications with changes in the intensive variables.  The free energy concept of Gibbs \cite{Gibbs_1875}, designed purely for changes in the extensive variables, is thus less versatile than the slightly older concept of Massieu \cite{Massieu_1869}, adopted by many well-known thermodynamicists \citep[e.g.][]{Planck_1932, Schrodinger_1946, Guggenheim_1967, Strong_H_1970, Callen_1985, Craig_1988} for this reason.  Concepts related to the free energy, such as essergy and exergy \cite{Keenan_1941, Keenan_1951, Rant_1956, Gaggioli_1962, Evans_1969, Moran_S_2006}, suffer from the same disadvantage (a further disadvantage is their need for an unchanging reference system).  In steady state systems, we could multiply $\phi^*_{st}$ by $kT/\theta \flow{V}$ to give the ``free power'', or by $k /\vect{\nabla} (T^{-1}) \theta \flow{V}$ to give a flux analogue, but neither choice makes much sense, and would involve the same awkwardness in the handling of intensive variables as does the free energy. If $\phi^*_{st}$ must have units, it is preferable to multiply it by $k/\theta \flow{V}$ to give it the same units as $\hat{\dot{\sigma}}$; being of opposite sign, we could call this (as would Gibbs \cite{Gibbs_1873b}) the ``available capacity for entropy production'' or (as would Brillouin \cite{Brillouin_1953}) the ``free negentropy production''. Provided that the dissipative processes are macroscopically reproducible, the fact that this quantity has entropy units (a non-conserved quantity) instead of energy units (a conserved quantity) will not affect its application.

\section{\label{Apx_laws}Steady State ``Laws''}  

Steady state analogues of the ``laws'' of equilibrium thermodynamics are readily obtainable from Jaynes' generic approach \cite{Tribus_1961b, Kapur_K_1992}\footnote{Tribus \cite{Tribus_1961b} would distinguish them as ``laws of thermostatics'' and ``laws of thermodynamics''.}.  Although unsurprising, they are included here for completeness.  Firstly, two control volumes which are joined (side-by-side) such that their fluxes are combined (i.e.\ $\langle \vect{j}_{m} \rangle_{tot} = \langle \vect{j}_{m} \rangle_1 + \langle \vect{j}_{m} \rangle_2 $ for $\langle \vect{j}_m \rangle \in \{ \langle \vect{j}_r \rangle \}$) will attain a steady state position of identical multipliers $\{\zeta_m\}$, and so will share common $m$th gradients; this is the zeroth law.  The first law is simply a definition of generalised heat and work for the flux of a conserved quantity, which at steady state is given by $d \langle \vect{j}_r \rangle = \delta \vect{q}_r + \delta \vect{w}_r$; thus $\delta \vect{q}_r$ gives a (spontaneous) change in the distribution of instantaneous fluxes, whilst $\delta \vect{w}_r$ represents a (recoverable) change in the instantaneous fluxes themselves. The second law can be stated in many ways (see \S\ref{Discussion}); in a state function sense, it is given by the generalised Clausius equality \eqref{eq:dHstar_st}. 
The third law -- arguably more of a definition of a convenient reference state -- can be stated as ``the steady state entropy $\mathfrak{H}^*_{st}$ approaches 
zero at the position of zero gradients'', i.e.\ when it relaxes to an equilibrium position. 
\section{\label{Apx_rels} Jaynes' Relations}  

In linear transport theory, it is common to employ the so-called ``Curie postulate'', in which the scalar, vector or tensor fluxes $j_r$, $\vect{j}_r$ or $\tens{j}_r$ are assumed to depend on forces or gradients of the same type $\Delta \lambda_r$, $\vect{\nabla} \lambda_r$ or $\tens{\nabla} \vect {\lambda}_r^\top$, but not on those of other types \cite{Curie_1894, deGroot_M_1962, Kreuzer_1981, Bird_etal_2006}.  This postulate cannot be assumed to apply in general.  In consequence, Jaynes'  relations \eqref{eq:diff1}-\eqref{eq:diff2mixed} and higher derivatives for scalar and each component of vector and tensor multipliers in a steady state system will be of generalised form:
\begin{gather}
\frac{\partial  \zeta_0^* }{\partial {\zeta}_{r{\imath \jmath}}}   = - \langle {{j}_{r{\imath \jmath}}} \rangle ,
\label{eq:diff1_st}
\\
\frac{\partial^2 \zeta_0^*}{\partial {\zeta}_{r{\imath \jmath}}^2}   = \langle {{{{j}_{r{\imath \jmath}}}}^2} \rangle- {\langle {{j}_{r{\imath \jmath}}} \rangle}^2   = - \frac {\partial \langle {{j}_{r{\imath \jmath}}} \rangle}{\partial {\zeta}_{r{\imath \jmath}}},
\label{eq:diff2_st} 
\\
\begin{split}
\frac{\partial^2 \zeta_0^*}{\partial {\zeta}_{m{\kappa \ell}} \, \partial {\zeta}_{r{\imath \jmath}}}    &= \langle {{{j}_{r{\imath \jmath}}}{{j}_{m{\kappa \ell}}}} \rangle - {\langle {{j}_{r{\imath \jmath}}} \rangle}{\langle {{j}_{m{\kappa \ell}}} \rangle}  \\ 
&= - \frac {\partial \langle {{j}_{r{\imath \jmath}}} \rangle}{\partial {\zeta}_{m{\kappa \ell}}}  = - \frac {\partial \langle {{j}_{m{\kappa \ell}}} \rangle}{\partial {\zeta}_{r{\imath \jmath}}}  
\end{split}
\label{eq:diff2mixed_st}
\\
\begin{split}
 &\frac{\partial^2 \zeta_0^*}{ \partial \zeta_{n \varphi \vartheta } \, \partial {\zeta}_{m{\kappa \ell}} \, \partial {\zeta}_{r{\imath \jmath}} }    
\\ 
&= - \frac {\partial \langle {{j}_{r{\imath \jmath}}} \rangle}{\partial \zeta_{n \varphi \vartheta } \partial {\zeta}_{m{\kappa \ell}}}  
= - \frac {\partial \langle {{j}_{m{\kappa \ell}}} \rangle}{\partial \zeta_{n \varphi \vartheta } \partial {\zeta}_{r{\imath \jmath}}}
= - \frac {\partial \langle {{j}_{n \varphi \vartheta } } \rangle}{ \partial {\zeta}_{r{\imath \jmath}} \partial {\zeta}_{m{\kappa \ell}} }
\end{split}
\label{eq:diff3mixed_st}
\\
\vdots
\notag
\end{gather}
where the expectation notation $\langle \cdot \rangle$ is reinstated; $\imath,\jmath, \kappa, \ell, \varphi, \vartheta \in \{x,y,z\}$ (with $\jmath, \ell, \vartheta$ redundant for vectors and all directions redundant for scalars). Each partial derivative is taken at constant other Lagrangian multipliers.  

Relations \eqref{eq:diff1_st}-\eqref{eq:diff2mixed_st} for the example system (Figure \ref{fig:CV}) are listed in Table \ref{table_st}. 
Analogous to those in Table \ref{table_eq}, the identities in Table \ref{table_st} apply only at steady state; i.e.\ they describe quasistatic transitions on the manifold of steady state positions. Note that these relations do not impose tensor symmetry; e.g.\ $\partial \langle \tau_{\imath\jmath} \rangle / \partial \zeta_{\tens{\tau}_{\jmath \imath}} =  \partial \langle \tau_{\jmath \imath} \rangle / \partial \zeta_{\tens{\tau}_{\imath\jmath}}$ does not imply $\langle \tau_{\imath\jmath} \rangle = \langle \tau_{\jmath\imath} \rangle$ unless $\zeta_{\tens{\tau}_{\imath \jmath}}=\zeta_{\tens{\tau}_{\jmath \imath}}$.

{Furthermore}, each flux can be expanded in a Taylor series about the zero-multiplier (equilibrium) position $\langle j_{r \imath \jmath} \rangle|_{\{\zeta_r=0\}} = 0$ (see Appendix \ref{Apx_laws}), yielding \cite{Callen_1985}:
\begin{gather}
\begin{split}
 \langle {{j}_{r{\imath \jmath}}} \rangle & = \sum_{m \kappa \ell}  L^{0}_{r \imath \jmath, m \kappa \ell} \;  {\zeta}_{m{\kappa \ell}} 
\\
& + 
\frac{1}{2!} \sum_{m \kappa \ell} \sum_{n \varphi \vartheta }  L^{0}_{r \imath \jmath, m \kappa \ell , n \varphi \vartheta } \;   {\zeta}_{m{\kappa \ell}} \;  \zeta_{n \varphi \vartheta }
+ ...
\end{split}
\label{eq:flux_exp}
\end{gather}
with the equilibrium-limit derivatives:
\begin{gather*}
L^{0}_{r \imath \jmath, m \kappa \ell}=  \frac {\partial \langle {{j}_{r{\imath \jmath}}} \rangle} {\partial {\zeta}_{m{\kappa \ell}}} \biggr|_{\{\zeta_r=0\}}
\\
L^{0}_{r \imath \jmath, m \kappa \ell , n \varphi \vartheta } = \frac {\partial^2 \langle {{j}_{r{\imath \jmath}}} \rangle} {\partial {\zeta}_{m{\kappa \ell}} \partial \zeta_{n \varphi \vartheta }} \biggr|_{\{\zeta_r=0\}}
\end{gather*}
Close to equilibrium, this can be approximated by discarding all but the first-order terms:
\begin{equation}
\begin{split}
& \langle {{j}_{r{\imath \jmath}}} \rangle \approx \sum_{m \kappa \ell}  L^{0}_{r \imath \jmath, m \kappa \ell} \;  {\zeta}_{m{\kappa \ell}} 
\end{split}
\label{eq:flux_exp_Onsager}
\end{equation}
{By the use of Jaynes' method, we} therefore recover the near-equilibrium linear regime \eqref{eq:flux_exp_Onsager} with, from \eqref{eq:diff2mixed_st}, the Onsager reciprocal relations $L^{0}_{r \imath \jmath, m \kappa \ell} = L^{0}_{m \kappa \ell, r \imath \jmath}$ \cite{Onsager_1931a, Onsager_1931b}.  For simple boundary conditions, these relations may apply to the overall system as well as at local scales. However, the analysis goes well beyond Onsager's, indicating that the reciprocal relations \eqref{eq:diff2mixed_st} and higher derivatives (\eqref{eq:diff3mixed_st} and onwards) also apply well away from equilibrium, at least when expressed in terms of the raw multipliers (see \S\ref{Imps}) at local scales, although they will not then enter into the expansions \eqref{eq:flux_exp}-\eqref{eq:flux_exp_Onsager}. {Of course, the expansion \eqref{eq:flux_exp} (hence \eqref{eq:flux_exp_Onsager}) will become more and more inaccurate with increasing distance from equilibrium}, even if successively higher order terms are included. From \eqref{eq:diff3mixed_st} and onwards, each set of higher order derivatives will also obey a symmetry relation, both near and far from equilibrium.  

\begin{table*}[t]
\caption{\label{table_st}Multiplier relations \eqref{eq:diff1_st}, variances \eqref{eq:diff2_st} and covariances \eqref{eq:diff2mixed_st} for the steady state thermodynamic system of Figure \ref{fig:CV}, at steady state (here $K=\tscale\Vscale / k$, whilst subscripts $b$ and $e$ respectively imply $b \neq c$ and $e \neq d$).  The parameters held constant in each partial derivative can be judged by context. 
}
\begin{tabular*}{\textwidth}{l}  
\hline\hline
\end{tabular*}
\begin{tabular*}{168pt}[t]{p{31pt} p{5pt} l}
\multicolumn{3}{l}{Multiplier relations}
\\
$ - \langle j_{Q\imath} \rangle$ &=&  $- \dfrac{\partial  \zeta_0^* }{ \partial \biggl[ K \dfrac{\partial}{\partial \imath} \biggl(\dfrac{1}{T} \biggr)  \biggr]}$    \vspace{5pt}
\\
$ - \langle j_{c \imath} \rangle$ &=& $\dfrac{\partial  \zeta_0^* }{\partial \biggl[ K  \biggl[ \dfrac{\partial}{\partial \imath} \biggl(\dfrac{\mu_c}{M_c T} \biggr) - \dfrac{g_{c\imath}}{T} \biggl] \biggr]}   $    \vspace{5pt}
\\
$ -  {\langle \tau_{\imath\jmath} \rangle}  $ &=& $\dfrac{\partial  \zeta_0^* }{\partial \biggl[ K  \dfrac{\partial}{\partial \imath} \biggl(\dfrac{v_\jmath}{T} \biggr)  \biggr]}   $    \vspace{5pt}
\\
$ - { \langle \hat{\dot{\xi}}_d \rangle}$ &=& $\dfrac{\partial  \zeta_0^* }{\partial \biggl(K  \dfrac{A_d}{T} \biggr)  }    $    \vspace{5pt}
\\
\hline
\multicolumn{3}{l}{Variances}
\\
$\var(j_{Q\imath})$ &=& $ \dfrac {\partial \langle j_{Q\imath} \rangle}{\partial \biggl[ K  \dfrac{\partial}{\partial \imath} \biggl(\dfrac{1}{T} \biggr)  \biggr]}$ \vspace{5pt}
\\
$\var(j_{c \imath})$ &=& $ - \dfrac {\partial \langle j_{c \imath} \rangle}{\partial \biggl[ K  \biggl[ \dfrac{\partial}{\partial \imath} \biggl(\dfrac{\mu_c}{M_c T} \biggr) - \dfrac{g_{c\imath}}{T} \biggl] \biggr]}$ \vspace{5pt}
\\
$\var(\tau_{\imath\jmath})$ &=& $ - \dfrac {\partial \langle \tau_{\imath\jmath} \rangle}{\partial \biggl[ K \dfrac{\partial}{\partial \imath} \biggl(\dfrac{v_\jmath}{T} \biggr)  \biggr]}   $ \vspace{5pt} 
\\
$\var(\hat{\dot{\xi}}_d)$ &=& $ - \dfrac {\partial \langle \hat{\dot{\xi}}_d \rangle}{\partial \biggl(K  \dfrac{A_d}{T} \biggr)  }$ \vspace{5pt}
\\
\hline
\end{tabular*} 
\begin{tabular*}{335pt}[t]{p{3pt} p{49pt} l r l l}
&\multicolumn{4}{l}{Covariances (Maxwell-like relations)}
\\&
$\cov(j_{Q\imath},j_{Q\jmath}) $ &=& $ \dfrac {\partial \langle j_{Q\imath} \rangle}{\partial \biggl[ K \dfrac{\partial}{\partial \jmath} \biggl(\dfrac{1}{T} \biggr)  \biggr]}
$ &=& $ \dfrac {\partial \langle j_{Q\jmath} \rangle}{\partial \biggl[ K \dfrac{\partial}{\partial \imath} \biggl(\dfrac{1}{T} \biggr)  \biggr]}   $  \vspace{5pt}
\\& 
$\cov(j_{Q\imath},j_{c\kappa}) $ &=& $ \dfrac {\partial \langle j_{Q\imath} \rangle}{\partial \biggl[ K \biggl[ \dfrac{\partial}{\partial \kappa} \biggl(\dfrac{\mu_c}{M_c T} \biggr) - \dfrac{g_{c\kappa}}{T} \biggr] \biggr]}   
$ &=& $-\dfrac {\partial \langle j_{c\kappa} \rangle}{\partial \biggl[K \dfrac{\partial}{\partial \imath} \biggl(\dfrac{1}{T} \biggr)  \biggr]}   $  \vspace{5pt} 
\\&
$\cov(j_{Q\imath}, \tau_{\kappa\ell}) $ &=& $\dfrac {\partial \langle j_{Q\imath} \rangle}{\partial \biggl[ K \dfrac{\partial}{\partial \kappa} \biggl(\dfrac{v_\ell}{T} \biggr)  \biggr]}
$ &=& $-\dfrac {\partial \langle \tau_{\kappa\ell} \rangle}{\partial \biggl[ K \dfrac{\partial}{\partial \imath} \biggl(\dfrac{1}{T} \biggr)  \biggr]}   $  \vspace{5pt}
\\&
$\cov(j_{Q\imath}, \hat{\dot{\xi}}_{d}) $ &=& $\dfrac {\partial \langle j_{Q\imath} \rangle}{ \partial \biggl(K \dfrac{A_d}{T} \biggr)  }
$ &=& $-\dfrac {\partial \langle \hat{\dot{\xi}}_{d} \rangle}{\partial \biggl[ K \dfrac{\partial}{\partial \imath} \biggl(\dfrac{1}{T} \biggr)  \biggr]}   $  \vspace{5pt}
\\&
$\cov(j_{c\imath},j_{c\jmath}) $ &=& $ -\dfrac {\partial \langle j_{c\imath} \rangle}{\partial \biggl[ K \biggl[ \dfrac{\partial}{\partial \jmath} \biggl(\dfrac{\mu_c}{M_c T} \biggr) - \dfrac{g_{c\jmath}}{T} \biggr] \biggr]}   
$ &=& $ -\dfrac {\partial \langle j_{c\jmath} \rangle}{\partial \biggl[ K \biggl[ \dfrac{\partial}{\partial \imath} \biggl(\dfrac{\mu_c}{M_c T} \biggr) - \dfrac{g_{c\imath}}{T} \biggr] \biggr]}   $  \vspace{5pt} 
\\&
$\cov(j_{c\imath},j_{b\kappa}) $ &=& $ -\dfrac {\partial \langle j_{c\imath} \rangle}{\partial \biggl[ K \biggl[ \dfrac{\partial}{\partial \kappa} \biggl(\dfrac{\mu_b}{M_b T} \biggr) - \dfrac{g_{b\kappa}}{T} \biggr] \biggr]}   
$ &=& $ -\dfrac {\partial \langle j_{b\kappa} \rangle}{\partial \biggl[ K \biggl[ \dfrac{\partial}{\partial \imath} \biggl(\dfrac{\mu_c}{M_c T} \biggr) - \dfrac{g_{c\imath}}{T} \biggr] \biggr]}   $  \vspace{5pt}
\\&
$\cov(j_{c\imath}, \tau_{\kappa\ell} ) $ &=& $ -\dfrac {\partial \langle j_{c\imath} \rangle}{\partial \biggl[ K \dfrac{\partial}{\partial \kappa} \biggl(\dfrac{v_\ell}{T} \biggr) \biggr]}   
$ &=& $ -\dfrac {\partial \langle \tau_{\kappa\ell} \rangle}{ \partial \biggl[ K \biggl[ \dfrac{\partial}{\partial \imath} \biggl(\dfrac{\mu_c}{M_c T} \biggr) - \dfrac{g_{c\imath}}{T} \biggr] \biggr]}   $  \vspace{5pt} 
\\&
$\cov(j_{c\imath}, \hat{\dot{\xi}}_d ) $ &=& $ -\dfrac {\partial \langle j_{c\imath} \rangle}{\partial \biggl(K \dfrac{A_d}{T} \biggr)  }  
$ &=& $ -\dfrac {\partial \langle \hat{\dot{\xi}}_d \rangle}{\partial \biggl[ K \biggl[ \dfrac{\partial}{\partial \imath} \biggl(\dfrac{\mu_c}{M_c T} \biggr) - \dfrac{g_{c\imath}}{T} \biggr] \biggr]}   $  \vspace{5pt} 
\\&
$\cov(\tau_{\imath\jmath}, \tau_{\kappa\ell} ) $ &=& $ -\dfrac {\partial \langle \tau_{\imath\jmath} \rangle}{ \partial \biggl[ K \dfrac{\partial}{\partial \kappa} \biggl(\dfrac{v_\ell}{T} \biggr)  \biggr]}
$ &=& $-\dfrac {\partial \langle \tau_{\kappa\ell} \rangle}{ \partial \biggl[ K \dfrac{\partial}{\partial \imath} \biggl(\dfrac{v_\jmath}{T} \biggr)  \biggr]} $  \vspace{5pt}
\\&
$\cov(\tau_{\imath\jmath}, \hat{\dot{\xi}}_d ) $ &=& $ -\dfrac {\partial \langle \tau_{\imath\jmath} \rangle}{ \partial \biggl(K \dfrac{A_d}{T} \biggr)  }
$ &=& $- \dfrac {\partial \langle \hat{\dot{\xi}}_d \rangle}{ \partial \biggl[ K \dfrac{\partial}{\partial \imath} \biggl(\dfrac{v_\jmath}{T} \biggr)  \biggr]} $  \vspace{5pt}
\\&
$\cov(\hat{\dot{\xi}}_d, \hat{\dot{\xi}}_e ) $ &=& $ -\dfrac {\partial \langle \hat{\dot{\xi}}_d \rangle}{ \partial \biggl(K \dfrac{A_e}{T} \biggr)  }
$ &=& $-\dfrac {\partial \langle \hat{\dot{\xi}}_e \rangle}{ \partial \biggl(K \dfrac{A_d}{T} \biggr)  } $  \vspace{5pt}
\\
\end{tabular*}
\begin{tabular*}{\textwidth}{l}  
\hline\hline
\end{tabular*}
\end{table*}


\pagebreak

{
\begin{longtable}{p{35pt} p{200pt}}
\caption[Nomenclature]{\label{table_nomen}{Nomenclature used in this study (ES =  equilibrium system; CV = control volume).}}\\
%
\hline\hline
\multicolumn{2}{l}%
{}\\
{\bf Symbol}&{\bf Meaning \vphantom{$^{T^T}$}}\\
\hline \hline
\endfirsthead
\caption[]{(continued)}\\
\hline\hline
\multicolumn{2}{l}%
{}\\
{\bf Symbol}&{\bf Meaning \vphantom{$^{T^T}$} }\\
\hline\hline
\endhead
\hline\hline
\endfoot
\hline\hline
\endlastfoot
%
\multicolumn{2}{l}{{\bf Roman symbols \vphantom{$^{T^T}$}} \vspace{2pt}}\\
$A_{d}$ &molar chemical affinity of reaction $d$ ($A_d<0$ for spontaneous forwards reaction)\\
$C$ &total number of chemical species\\
$D$ &Kullback-Leibler cross-entropy function (generic) \\
$f_{ri}$  &value of property $f_r$ in the $i$th category (generic) \\
$\langle f_{r} \rangle$  &mathematical expectation of property $f_r$ (generic) \\
${\flow{F}_r}$ &flow rate of property $f_r$ (CV)\\
${\flow{F}_{ri}}$ &instantaneous flow rate of $f_r$ (CV)\\
${\vect g}_c$ &specific body force on chemical species $c$ (CV)\\
$\mathfrak{H}$ &entropy function (generic) \\
${\vect j}_{\flow{N}_c}$ &instantaneous mass flux of chemical species $c$ through element (relative to ${\vect v}$) (CV)\\
$\vect{j}_{s,tot}$ &total thermodynamic entropy flux (CV)\\
${\vect j}_c$ &mean mass flux of chemical species $c$ through element (relative to ${\vect v}$) (CV)\\
${\vect j}_Q$ &mean local heat flux through element (CV)\\
${\vect j}_{Q,\flow{I}}$ &instantaneous value of heat flux through element (CV)\\
$J$ &free energy for system with $\langle U \rangle$, $\langle V \rangle$ and $\langle n_c \rangle$ constraints (ES)\\
$k$ &Boltzmann constant\\
$m_c$ &specific (per unit fluid mass) moles of particles of chemical species $c$ (CV)\\
$M_c$ &molecular mass of chemical species $c$\\
$\langle n_c \rangle$  &mean moles of particles of chemical species $c$ (ES)\\
$n_{N_c}$ &moles of particles of chemical species $c$ (ES)\\
$N$  &total number of particles or entities\\
$p_i$  &probability of the $i$th category (generic)\\
$p_{i,j,\{N_c\}}$  &joint probability of each category in ES\\
$P$ &absolute pressure ($P>0$ for compression)\\
$q_i$ &source (prior) probability of $i$th category (generic)\\
$\delta \vect{q}_r$ &increment in ``generalised heat'' flux associated with $r$th constraint (CV)\\
$\delta Q$ &increment in actual (thermodynamic) heat (ES)\\
$\delta Q_r$ &increment in ``generalised heat'' associated with $r$th constraint (generic, ES)\\
$R$	&total number of moment constraints\\
$s$  &number of categories (generic, ES); specific thermodynamic entropy (CV)\\
$S$ &thermodynamic entropy (ES)\\
$T$ &absolute temperature\\
$t$ &time\\
$u$ &specific (per unit fluid mass) internal energy (CV)\\
$\langle U \rangle$ &mean internal energy (ES)\\
$U_i$ &internal energy level (ES)\\
${\vect v}$ &mass-average velocity through element (CV)  \\
$V$ &volume\\
$\langle V \rangle$ &mean volume (ES)\\
$V_j$ &volume element ``level'' (ES)\\
$\flow{V}$	&characteristic volume length scale (CV)\\
$\delta \vect{w}_r$ &increment in ``generalised work'' flux associated with $r$th constraint (CV)\\
$\delta W$ &increment in actual (thermodynamic) work (ES)\\
$\delta W_r$ &increment in ``generalised work'' associated with $r$th constraint (generic, ES)\\
$Z_q^*$ &partition function (generic)\\
$\flow{Z}$	&partition function (CV)\\
\hline
\multicolumn{2}{l}{{\bf Greek symbols \vphantom{$^{T^T}$}} \vspace{2pt}}\\
$\alpha_{i,j,\{N_c\}}^*$ &defined in \eqref{eq:pstar2_i_thermo} (ES)\\
$\beta_{\vecti}^*$ &defined in \eqref{eq:pi_star} (CV)\\
${\tens {\delta}}$ &Kronecker delta tensor\\
$\zeta_0^*$ &Massieu function = Lagrangian multiplier for the natural constraint (CV)\\
$\zeta_r$  &Lagrangian multiplier for the $r$th constraint (CV)\\
$\theta$	&characteristic time length scale (CV)\\
$\lambda_0^*$ &Massieu function = Lagrangian multiplier for the natural constraint (generic, ES)\\
$\lambda_r$  &Lagrangian multiplier for the $r$th constraint (generic, ES)\\
$\mu_c$ &molar chemical potential of chemical species $c$\\
$\nu_{cd}$ &stoichiometric coefficient of chemical species $c$ in the $d$th reaction ($\nu_{cd}>0$ for a product)\\
$\xi_{d}$ &extent of chemical reaction $d$\\
$\hat{\dot{\xi}}_{d}$ &mean molar rate per unit volume of chemical reaction $d$ (CV)\\
$\hat{\dot{\xi}}_{\flow{L}_{d}}$ &instantaneous molar rate per unit volume of chemical reaction $d$ (CV)\\
$\Xi$ &grand partition function (ES)\\
$\pi_{\vecti}$ &joint probability of each category in CV\\
${\tens {\Pi}}$ &net or molecular stress tensor in element (CV)\\
$\rho$ &fluid density\\
$\sigma$ &entropy produced and exported from system (ES)\\
$\dot{\sigma}$ &rate of production of thermodynamic entropy in control volume (CV)\\
$\hat{\dot{\sigma}}$ &rate of production, per unit volume, of thermodynamic entropy in element (CV)\\
${\tens {\tau}}$ &mean viscous stress tensor in element (wherein $\tau_{ij}>0$ for compression) (CV)\\
${{\tens{\tau}}}_{\flow{J}}$ &instantaneous value of viscous stress tensor in element (CV)\\
$\psi$ &negative Planck potential for system with $\langle U \rangle$, $\langle V \rangle$ and $\langle n_c \rangle$ constraints (ES) \\
$\phi$ &potential function (generic)\\
\hline
\multicolumn{2}{l}{{\bf Superscripts \vphantom{$^{T^T}$}} \vspace{2pt}}\\
$^*$  &at a stationary position\\
\hline
\multicolumn{2}{l}{{\bf Subscripts \vphantom{$^{T^T}$}} \vspace{2pt}}\\
$c$ &chemical species index (ES); pertaining to the ${\vect j}_c$ constraint (CV)\\
$CV$ &control volume (CV)\\
$d$ &chemical reaction index (ES); pertaining to the $\hat{\dot{\xi}}_{d}$ constraint (CV)\\
$eq$ &an equilibrium (quantity controlled) system (ES)\\
$i$ &index of categories (generic); index of energy levels (ES)\\
$\imath, \jmath$	&Cartesian coordinate indices\\
$\flow{I}$ &index of instantaneous heat flux (CV)\\
${\vecti}$ &joint index of heat flux, mass fluxes of chemical species $c$, viscous stress tensor and rates of chemical reactions $d$ in element (CV)\\
$in$ &into control volume (CV)\\
$j$ &index of volume elements (ES)\\
$\flow{J}$ &index of instantaneous viscous stress tensor (CV)\\
$\kappa, \ell$	&Cartesian coordinate indices\\
$\flow{L}_{d}$ &index of instantaneous rate per volume of chemical reaction $d$ (CV)\\
$max$ &maximum\\
$min$ &minimum\\
$m$ &index of the $m$th constraint (generic)\\
$n$ &index of the $n$th constraint (generic)\\
$n_c$ &pertaining to the $\langle n_c \rangle$ constraint (ES)\\
$N_c$ &number of particles of chemical species $c$ (ES)\\
$\flow{N}_c$ &index of instantaneous mass flux of chemical species $c$ (CV)\\
$opt$ &optimum\\
$out$ &out of control volume (CV)\\
$Q$ &pertaining to the ${\vect j}_Q$ constraint (CV)\\
$r$ &index of the $r$th constraint (generic)\\
$st$ &a steady state (flow-controlled) system (CV)\\
$U$ &pertaining to $\langle U \rangle$ constraint (ES)\\
$V$ &pertaining to $\langle V \rangle$ constraint (ES)\\
${\tens {\tau}}$ &pertaining to the ${\tens {\tau}}$ constraint (CV)\\
$\varphi, \vartheta$	&Cartesian coordinate indices\\
\hline
\multicolumn{2}{l}{{\bf Mathematical symbolism \vphantom{$^{T^T}$}} \vspace{2pt}}\\
$\langle x \rangle$  &mathematical expectation of $x$ \\
$dx$ &total differential of $x$\\
$\delta x$ &path-dependent differential of $x$\\
$\vect{\nabla}$ &Cartesian gradient (vector) operator\\
$\vect{\nabla} \cdot$ &Cartesian divergence (scalar) operator\\
$\var(a)$ &variance of $a$\\
$\cov(a,b)$ &covariance of $a$ and $b$\\
$\vect{a}$ &vector quantity\\
$\vect{a}^\top$ &transpose of $\vect{a}$\\
$\vect{a} \cdot \vect{b}$ &vector scalar product (dot product)\\
$\vect{a} \vect{b}^\top$ &vector dyadic product\\
$\tens{A}$ &tensor quantity\\
$\tens{A} : \tens{B}$ &tensor scalar product\\
$\dot{x}$ &rate of production of $x$ within control volume (CV)\\
$\hat{\dot{x}}$ &rate of production per unit volume of $x$ within control volume (CV)\\
$x|_{a,b,...}$ &quantity $x$ taken at fixed values of $a, b, ...$\\
$\sum\nolimits_a^b x$ &sum of $x$ from $a$ to $b$\\
%
\end{longtable}
}

\pagebreak
\newpage


\begin{thebibliography}{150}
\bibitem{Paltridge_1975}G.W. Paltridge,\ti{ Global dynamics and climate - a system of minimum entropy exchange,} Quart. J. Royal Meteorol. Soc. {\vol 101}, 475\ti{-484} (1975).
\bibitem{Paltridge_1978}G.W. Paltridge,\ti{ The steady-state format of global climate,} Quart. J. Royal Meteorol. Soc. {\vol 104}, 927\ti{-945} (1978).
\bibitem{Paltridge_1981}G.W. Paltridge,\ti{ Thermodynamic dissipation and the global climate system,} Quart. J. Royal Meteorol. Soc. {\vol 107}, 531\ti{-547} (1981).
\bibitem{Ozawa_Ohmura_1997}H. Ozawa, A. Ohmura,\ti{ Thermodynamics of a global-mean state of the atmosphere -- a state of maximum entropy increase,} J. Clim. {\vol 10}, 441\ti{-445} (1997).
\bibitem{Paltridge_2001}G.W. Paltridge,\ti{ A physical basis for a maximum of thermodynamic dissipation of the climate system,} Quart. J. Royal Meteorol. Soc. {\vol 127}, 305\ti{-313} (2001).
\bibitem{Shimokawa_O_2001} S. Shimokawa, H. Ozawa,\ti{ On the thermodynamics of the oceanic general circulation: entropy increase rate of an open dissipative system and its surroundings,} Tellus {\vol 53A}, 266\ti{-277} (2001).
\bibitem{Shimokawa_O_2002} S. Shimokawa, H. Ozawa,\ti{ On the thermodynamics of the oceanic general circulation: Irreversible transition to a state with higher rate of entropy production,} Quart. J. Royal Meteorol. Soc. {\vol 128}, 2115\ti{-2128} (2002).
\bibitem{Kleidon_etal_2003} A. Kleidon, K. Fraedrich, T. Kunz, F. Lunkeit,\ti{ The atmospheric circulation and states of maximum entropy production,} Geophys. Res. Lett. {\vol 30}(23), article 2223 (2003).
\bibitem{Kleidon_L_book_2005} A. Kleidon, R.D. Lorenz (eds.) Non-equilibrium Thermodynamics and the Production of Entropy: Life, Earth and Beyond, Springer Verlag, Heidelberg, 2005.
 \bibitem{Kleidon_L_art_2005} A. Kleidon, R.D. Lorenz,\ti{ Entropy production by Earth system processes,} in A. Kleidon, R.D. Lorenz (eds.) Non-equilibrium Thermodynamics and the Production of Entropy: Life, Earth and Beyond, Springer Verlag, Heidelberg, 2005, 1\ti{-20}.
\bibitem{Ozawa_etal_2003}H. Ozawa, A. Ohmura, R.D. Lorenz, T. Pujol,\ti{ The second law of thermodynamics and the global climate system: A review of the maximum entropy production principle,} Rev. Geophys. {\vol 41}, article 4 (2003). 
\bibitem{Davis_2008}B. Davis,\ti{ MEP and the latitudinal temperature gradient,} MEP in the Earth System workshop, Max-Planck-Institut f\"ur Biogeochemie, Jena, Germany, 6-9 May 2008.
\bibitem{Paltridge_etal_2007} G.W. Paltridge, G.D. Farquhar, M. Cuntz,\ti{ Maximum entropy production, cloud feedback and climate change,} Geophys. Res. Lett. {\vol 34}, L14708 (2007).
\bibitem{Lorenz_etal_2001}R.D. Lorenz, J.I. Lunine, P.G. Withers, C.P. McKay,\ti{ Titan, Mars and Earth: Entropy production by latitudinal heat transport,} Geophys. Res. Lett. {\vol 28}, 415\ti{-418} (2001).
\bibitem{Goody_2007} R. Goody,\ti{ Maximum entropy production in climate theory,} J. Atm. Sci. {\vol 64}, 2735\ti{-2739} (2007).
\bibitem{Benard_1901}H. B\'enard,\ti{ Les tourbillons cellulaires dans une nappe liquide transportant de la chaleur par convection en r\'egime permanent,} Annales de Chimie et de Physique {\vol 23}, 62\ti{-144} (1901).
\bibitem{Ozawa_etal_2001}H. Ozawa, S. Shikokawa, H. Sakuma,\ti{ Thermodynamics of fluid turbulence: A unified approach to the maximum transport properties,} Phys. Rev. E {\vol 64}, article 026303 (2001). 
\bibitem{Vanyo_Paltridge_1981}J.P. Vanyo, G.W. Paltridge,\ti{ A model for energy dissipation at the mantle-core boundary,} Geophys. J. Royal Astron. Soc. {\vol 66}, 677\ti{-690} (1981).
\bibitem{Lorenz_2001b}R.D. Lorenz,\ti{ Of course Ganymede and Callisto have oceans: Application of a principle of maximum entropy production to icy satellite convection,} Proc. Lunar Planet Sci. Conf. {\vol 32}, abstract 1160 (2001).
\bibitem{Lorenz_2002a}R.D. Lorenz,\ti{ Planets, life and the production of entropy,} Int. J. Astrobiol. {\vol 1}, 3\ti{-13} (2002).
\bibitem{Kleidon_2004} A. Kleidon,\ti{ Beyond Gaia: thermodynamics of life and Earth system functioning,} Climatic Change {\vol 66}, 271\ti{-319} (2004).
\bibitem{Kleidon_S_2008} A. Kleidon, S. Schymanski,\ti{ Thermodynamics and optimality of the water budget on land: a review,} Geophys. Res. Lett. {\vol 35}, L20404 (2008).
\bibitem{Kleidon_F_L_2007} A. Kleidon, K. Fraedrich, C. Low,\ti{ Multiple steady-states in the terrestrial atmosphere-biosphere system: a result of a discrete vegetation classification?,} Biogeosciences {\vol 4}, 707\ti{-714} (2007).
\bibitem{Meysman_B_2007} F.J.R. Meysman, S. Bruers,\ti{ A thermodynamic perspective on food webs: Quantifying entropy production within detrital-based ecosystems,} J. Theor. Biol. {\vol 249}, 124\ti{-139} (2007).
\bibitem{Bruers_M_2007} S. Bruers, F.J.R. Meysman,\ti{ A useful correspondence between fluid convection and ecosystem operation,} {\it arXiv:0708.0091v1}, 2007.
\bibitem{Paulus_G_2004} D.M. Paulus, R.A. Gaggioli,\ti{ Some observations of entropy extrema in fluid flow,} Energy {\vol 29}, 2487\ti{-2500} (2004).
\bibitem{Zupanovic_etal_2004}P. \v{Z}upanovi\'c, D. Jureti\'c, S. Botri\'c,\ti{ KirchhoffÕs loop law and the maximum entropy production principle,} Phys. Rev. E {\vol 70}, article 056108 (2004).
\bibitem{Botric_etal_2005}S. Botri\'c, P. \v{Z}upanovi\'c, D. Jureti\'c,\ti{ Is the stationary current distribution in a linear planar electric network determined by the principle of maximum entropy production?,} Croatica Chemica Acta {\vol 78}(2), 181\ti{-184} (2005). 
\bibitem{Christen_2006} T. Christen,\ti{ Application of the maximum entropy production principle to electrical systems,} J. Phys. D: Appl. Phys. {\vol 39}, 4497\ti{-4503} (2006).
\bibitem{Bruers_etal_2007a} S. Bruers, C. Maes, K. Neto\v{c}n\'{y},\ti{ On the validity of entropy production principles for linear electrical circuits,} J. Stat. Phys. {\vol 129}, 725\ti{-740} (2007).
\bibitem{Christen_2007a} T. Christen,\ti{ A maximum entropy production model for Teflon ablation by arc radiation,} J. Phys. D: Appl. Phys. {\vol 40}, 5719\ti{-5722} (2007).
\bibitem{Yoshida_M_2008}Z. Yoshida, S.M. Mahajan,\ti{ ``Maximum'' entropy production in self-organized plasma boundary layer: A thermodynamic discussion about turbulent heat transport,} Physics of Plasmas {\vol 15}, article 032307 (2008).
\bibitem{Martyushev_A_2003} L.M. Martyushev, E.G. Axelrod,\ti{ From dendrites and S-shaped growth curves to the maximum entropy production principle,} JETP Letters {\vol 78}(8), 476\ti{-479} (2003).
\bibitem{Zupanovic_J_2004}P. \v{Z}upanovi\'c, D. Jureti\'c,\ti{ The chemical cycle kinetics close to the equilibrium state and electrical circuit analogy,} Croatica Chemica Acta {\vol 77}(4), 561\ti{-571} (2004). 
\bibitem{Christen_2007b} T. Christen,\ti{ Modelling diffusion in nonuniform solids using entropy production rate,} J. Phys. D: Appl. Phys. {\vol 40}, 5723\ti{-5726} (2007).
\bibitem{Juretic_Z_2003} D. Jureti\'c, P. \v{Z}upanovi\'c,\ti{ Photosynthetic models with maximum entropy production in irreversible charge transfer steps,} Computational Biology and Chemistry {\vol 27}, 541\ti{-553} (2003).
\bibitem{Dewar_etal_2006} R.C. Dewar, D. Jureti\'c, P. \v{Z}upanovi\'c,\ti{ The functional design of the rotary enzyme ATP synthase is consistent with maximum entropy production,} Chem. Phys. Lett. {\vol 430}, 177\ti{-182} (2006).
\bibitem{Jenkins_2005} A.D. Jenkins,\ti{ Thermodynamics and economics,} {\it arXiv:cond-mat/0503308v1}, 2005.
\bibitem{Martyushev_S_2006} L.M. Martyushev, V.D. Seleznev,\ti{ Maximum entropy production principle in physics, chemistry and biology,} Physics Reports {\vol 426}, 1\ti{-45} (2006).
\bibitem{Bruers_2007c} S. Bruers,\ti{ Classification and discussion of macroscopic entropy production principles,} {\it arXiv:cond-mat/0604482v3}, 2007.  
\bibitem{Lovelock_1988} J.E. Lovelock, Ages of Gaia, Norton, NY, 1988.
\bibitem{Karnani_A_2009} M. Karnani, A. Annila,\ti{ Gaia again,} Biosystems {\vol 95}(1), 82\ti{-87} (2009).
\bibitem{Dewar_2003}R.C. Dewar,\ti{ Information theory explanation of the fluctuation theorem, maximum entropy production and self-organized criticality in non-equilibrium stationary states,} J. Phys. A: Math. Gen. {\vol 36}, 631\ti{-641} (2003).
\bibitem{Dewar_2005}R.C. Dewar,\ti{ Maximum entropy production and the fluctuation theorem,} J. Phys. A: Math. Gen. {\vol 38}, L371\ti{-L381} (2005). 
\bibitem{Jaynes_1957} E.T. Jaynes,\ti{ Information theory and statistical mechanics,} Phys. Rev., {\vol 106}, 620\ti{-630} (1957).
\bibitem{Jaynes_1957b} E.T. Jaynes,\ti{ Information theory and statistical mechanics II,} Phys. Rev. {\vol 108}, 171\ti{-190} (1957).
\bibitem{Jaynes_1963} E.T. Jaynes,\ti{ Information theory and statistical mechanics,} {\it in} Ford, K.W. (ed), Brandeis University Summer Institute, Lectures in Theoretical Physics, Vol. 3: Statistical Physics, Benjamin-Cummings Publ. Co., 1963, 181\ti{-218}. 
\bibitem{Tribus_1961a} M. Tribus,\ti{ Information theory as the basis for thermostatics and thermodynamics,} J. Appl. Mech., Trans. ASME, {\vol 28}, 1\ti{-8} (1961).
\bibitem{Tribus_1961b} M. Tribus, Thermostatics and Thermodynamics, D. Van Nostrand Co. Inc., Princeton, NJ, 1961.
\bibitem{Kapur_K_1992} J.N. Kapur, H.K. Kesevan, Entropy Optimization Principles with Applications, Academic Press, Inc., Boston, MA, 1992.
\bibitem{Jaynes_2003} E.T. Jaynes (G.L. Bretthorst, ed.) Probability Theory: The Logic of Science, Cambridge U.P., Cambridge, 2003.
\bibitem{Bruers_2007d} S. Bruers,\ti{ A discussion on maximum entropy production and information theory,} {\it arXiv:0705.3226v1} (2007); J. Phys. A - Math. Theor. {\vol 40}(27) 7441\ti{-7450} (2007).
\bibitem{Grinstein_L_2007} G. Grinstein, R. Linsker,\ti{ Comments on a derivation and application of the `maximum entropy production' principle,} J. Phys. A: Math. Theor. {\vol 40}, 9717\ti{-9720} (2007).
\bibitem{Onsager_1931a} L. Onsager,\ti{ Reciprocal relations in irreversible processes I,} Phys. Rev. {\vol 37}, 405\ti{-426} (1931).
\bibitem{Onsager_1931b} L. Onsager,\ti{ Reciprocal relations in irreversible processes II,} Phys. Rev. {\vol 38}, 2265\ti{-2279} (1931).
\bibitem{Attard_2006a} P. Attard,\ti{ Statistical mechanical theory for steady state systems. VI. Variational principles,} J. Chem. Phys. {\vol 125}, article 214502 (2006).
\bibitem{Attard_2006b} P. Attard,\ti{ Theory for non-equilibrium statistical mechanics,} Phys. Chem. Chem. Phys. {\vol 8}, 3585\ti{-3611} (2006).
\bibitem{Beretta_2001}G.P. Beretta,\ti{ Maximal-entropy-production-rate nonlinear quantum dynamics compatible with second law, reciprocity, fluctuation-dissipation, and time-energy uncertainty relations,} {\it arXiv:quant-ph/0112046v1} (2001).
\bibitem{Gyft_Beretta_2005}E.P. Gyftopoulos, G.P. Beretta,\ti{ What is the second law of thermodynamics and are there any limits to its validity?,} {\it arXiv:quant-ph/0507187} (2005).
\bibitem{Beretta_2006}G.P. Beretta,\ti{ Nonlinear model dynamics for closed-system, constrained, maximal-entropy-generation relaxation by energy redistribution,} Phys. Rev. E {\vol 73}, 026113 (2006).
\bibitem{Beretta_2008}G.P. Beretta,\ti{ Modeling non-equilibrium dynamics of a discrete probability distribution: general rate equation for maximal entropy generation in a maximum-entropy landscape with time-dependent constraints,} Entropy {\vol 10}, 160-182 (2008).
\bibitem{Zupanovic_etal_2006}P. \v{Z}upanovi\'c, S. Botri\'c, D. Jureti\'c,\ti{ Relaxation processes, MaxEnt formalism and Einstein's formula for the probability of fluctuations,} Croatica Chemica Acta {\vol 79}(3), 335\ti{-338} (2006). 
\bibitem{Martyushev_2007} L.M. Martyushev,\ti{ Do nonequlibrium processes have features in common?,} {\it arXiv:0709.0152v1}, 2007.
\bibitem{Shannon_1948} C.E. Shannon,\ti{ A mathematical theory of communication,} Bell Sys. Tech. J. {\vol 27}, 379\ti{-423}; 623\ti{-659} (1948).
\bibitem{Shore_J_1980} J.E. Shore, R.W. Johnson,\ti{ Axiomatic derivation of the principle of maximum entropy and the principle of minimum cross-entropy,} IEEE Trans. Information Theory {\vol IT-26}(1), 26\ti{-37} (1980).
\bibitem{Boltzmann_1877} L. Boltzmann,\ti{ \"Uber die Beziehung zwischen dem zweiten Hauptsatze dewr mechanischen W\"armetheorie und der Wahrscheinlichkeitsrechnung, respective den S\"atzen \"uber das W\"armegleichgewicht,} Wien. Ber. {\vol 76}, 373\ti{-435} (1877); English transl.: J. Le Roux (2002) 1\ti{-63} {\it http://www.essi.fr/$\sim$leroux/}.
\bibitem{Planck_1901} M. Planck,\ti{ \"Uber das gesetz der Energieverteilung im Normalspektrum,} Annalen der Physik {\vol 4}, 553\ti{-563} (1901). 
\bibitem{Vincze_1972} I. Vincze,\ti{ On the maximum probability principle in statistical physics,} Progress in Statistics 
{\vol 2}, 869\ti{-895} (1974).
\bibitem{Grendar_G_2001}M. Grendar, M. Grendar,\ti{ What is the question that MaxEnt answers? A probabilistic interpretation,} {\it in} A. Mohammad-Djafari (ed.) 
MaxEnt 2000, Gif-sur-Yvette, France, 8-13 July 2000, AIP Conf. Proc. {\vol 568}, 83\ti{-94} (2001).
\bibitem{Niven_CIT}R.K. Niven,\ti{ Combinatorial information theory: I. Philosophical basis of cross-entropy and entropy,} {\it http://arxiv.org/abs/cond-mat/0512017 v5}, 2007.
\bibitem{Niven_MaxEnt07} R.K. Niven,\ti{ Origins of the combinatorial basis of entropy,} {\it in} K.H. Knuth, A. Caticha, J.L. Center, A. Giffon, C.C. Rodr\' guez (eds), MaxEnt 2007, Saratoga Springs, NY, 8-13 July 2007, AIP Conf. Proc. {\vol 954}, 133\ti{-142} (2007).
\bibitem{Stirling_1730} J. Stirling, Methodus Differentialis: Sive Tractatus de Summatione et Interpolatione Serierum Infinitarum, Gul. Bowyer, London, 1730, Propositio XXVII, 135\ti{-139}.
\bibitem{Sanov_1957}I. N. Sanov,\ti{ On the probability of large deviations of random variables,} Mat. Sbornik {\vol 42}, 11\ti{-44} (1957) (Russian).
\bibitem{Kullback_L_1951} S. Kullback, R.A. Leibler,\ti{ On information and sufficiency,} Annals Math. Stat. {\vol 22}, 79\ti{-86} (1951).
\bibitem{Kullback_1959} S. Kullback, Information Theory and Statistics, John Wiley, NY, 1959.
\bibitem{Massieu_1869} M. Massieu,\ti{ Thermodynamique - Sur les fonctions caract\'eristiques des divers fluides,} Comptes Rendus {\vol 69}, 858\ti{-862}; 1057\ti{-1061} (1869).
\bibitem{Callen_1985} H.B. Callen, Thermodynamics and an Introduction to Thermostatistics, 2nd ed., John Wiley, NY, 1985.
\bibitem{Gibbs_1902} J.W. Gibbs, Elementary Principles of Statistical Mechanics, Dover Publ., NY, 1902.
\bibitem{Einstein_1902} A. Einstein,\ti{ Kinetische Theorie des W\"armegleichgewichtes und des zweiten Hauptsatzes der Thermodynamik,} 
Annalen der Physik {\vol 9}, 417\ti{-433} (1902).
\bibitem{Einstein_1903} A. Einstein,\ti{ Eine Theorie der Grundlagen der Thermodynamik,} 
Annalen der Physik {\vol 11}, 170\ti{-187} (1903).
%
\bibitem{Clausius_1865} R. Clausius,\ti{ \"Uber verschiedene f\"ur die Anwendung bequeme Formen der Hauptgleichungen der mechanischen WŠrmetheorie,} Poggendorfs Annalen {\vol 125}, 335\ti{-400} (1865); English transl.: R.B. Lindsay {\it in} J. Kestin (ed.) {The Second Law of Thermodynamics}, Dowden, Hutchinson \& Ross, PA (1976) 162\ti{-193}.
\bibitem{Gibbs_1873b} J.W. Gibbs,\ti{ A method of graphical representation of the thermodynamic properties of substances by means of surfaces,} Trans. Connecticut Acad. {\vol 2}, 382\ti{-404} (1873). 
\bibitem{Gibbs_1875} J.W. Gibbs,\ti{ On the equilibrium of heterogeneous substances,} Trans. Connecticut Acad. {\vol 3}, 108\ti{-248} (1875-1876); 343\ti{-524} (1877-1878) .
\bibitem{Keenan_1941} J.H. Keenan, Thermodynamics, John Wiley, NY, 1941.
\bibitem{Keenan_1951} J.H. Keenan,\ti{ Availability and irreversibility in thermodynamics,} Brit. J. Appl. Phys. {\vol 2}, 183\ti{-193} (1951).
\bibitem{Rant_1956} Z. Rant,\ti{ Exergie, ein neues Wort fur, technische Arbeitsfahigkeit,} Forschung im Ingenieurwesen {\vol 22}(1), 36\ti{-37} (1956).
\bibitem{Gaggioli_1962} R.A. Gaggioli,\ti{ The concepts of thermodynamic friction, thermal available energy, chemical available energy and thermal energy,} Chem. Eng. Sci. {\vol 17}, 523\ti{-530} (1962).
\bibitem{Evans_1969} R.B. Evans,\ti{ A Proof that Essergy is the Only Consistent Measure of Potential Work (for Chemical Substances),} PhD thesis, Dartmouth College, NH, 1969 ({\it unpub.}).
\bibitem{Moran_S_2006} M.J. Moran, H.N. Shapiro, Fundamentals of Engineering Thermodynamics, 5th ed., John Wiley, NY, 2006.
\bibitem{Gaggioli_etal_2002a} R.A. Gaggioli, D.H. Richardson, A.J. Bowman,\ti{ Available energy -- Part I: Gibbs revisited,} J. Energy Resources Technol., ASCE {\vol 124}, 105\ti{-109} (2002).
\bibitem{Gaggioli_etal_2002b} R.A. Gaggioli, D.M. Paulus Jr,\ti{ Available energy -- Part II: Gibbs extended,} J. Energy Resources Technol., ASCE {\vol 124}, 110\ti{-115} (2002).
\bibitem{Prigogine_1967} I. Prigogine, Introduction to Thermodynamics of Irreversible Processes, 3rd ed., Interscience Publ., NY, 1967.
\bibitem{Kondepudi_P_1998} D. Kondepudi, I. Prigogine, Modern Thermodynamics: from Heat Engines to Dissipative Structures, John Wiley \& Sons, Chichester, UK, 1998.
\bibitem{Planck_1922} M. Planck, Treatise on Thermodynamics, Engl. transl., 3rd ed., Dover Publications, NY, 1945.  
\bibitem{Planck_1932} M. Planck, Introduction to Theoretical Physics, Vol. V: Theory of Heat, Engl. transl. H.L. Brose, Macmillan \& Co., Ltd, 1932.
\bibitem{Fermi_1936} E. Fermi, Thermodynamics, Dover Publ., NY, 1956.
\bibitem{Strong_H_1970} L.E. Strong, H.F. Halliwell,\ti{ An alternative to free energy for undergraduate instruction,} J. Chem. Ed. {\vol 47}(5), 347\ti{-352} (1970).
\bibitem{Craig_1988} N.C. Craig,\ti{ Entropy analysis of four familiar processes,} J. Chem. Ed. {\vol 65}(9), 760\ti{-764} (1988).
\bibitem{Schrodinger_1946} E. Schr\"{o}dinger, Statistical Thermodynamics, Dover Publ., NY, 1989.
\bibitem{Guggenheim_1967} E.A. Guggenheim, Thermodynamics: An Advanced Treatment for Chemists and Physicists, North-Holland Publ. Co., Amsterdam, 1967.
\bibitem{Maxwell_1888} J.C. Maxwell, Theory of Heat, 9th ed., Dover Publ. NY, 2001.
\bibitem{Lanczos_1966} C. Lanczos, The Variational Principles of Mechanics, 3rd ed., University of Toronto Press, Toronto, 1966.
\bibitem{Boltzmann_1872} L. Boltzmann,\ti{ Weitere Studien \"{u}ber das W\"{a}rmegleichgewicht unter Gasmolek\"{u}len,} Sitzungsberichte Akad. Wissenschaften {\vol 66}, 275\ti{-370} (1872); Engl. transl., in S.G. Brush (ed.), Kinetic Theory, {\vol 2}, Pergamon Press, Oxford, 1966, 88\ti{-174}.
%
\bibitem{Grendar_G_2004a} M. Grendar, M. Grendar,\ti{ Maximum entropy method with non-linear moment constraints: challenges,} in G. Erickson and Y. Zhai (eds.) 
MaxEnt 2003, Jackson Hole, Wyoming, USA, 3-8 August 2003, AIP Conf. Proc. {\vol 707}, 97\ti{-109} (2004).
\bibitem{deGroot_M_1962} S.R. de Groot, P. Mazur, Non-Equilibrium Thermodynamics, Dover Publications, NY, 1984.
\bibitem{Kreuzer_1981} H.J. Kreuzer, Nonequilibrium Thermodynamics and its Statistical Foundations, Clarendon Press, Oxford, 1981.
\bibitem{Bird_etal_2006} R.B. Bird, W.E. Stewart, E.N. Lightfoot, Transport Phenomena, 2nd ed., John Wiley \& Sons, NY, 2002.
\bibitem{White_2005} F.M. White, Viscous Fluid Flow, 3rd ed., McGraw-Hill, NY (2005).
%
\bibitem{Jou_etal_1993} D. Jou, J. Casas-V\'azquez, G. Lebon, Extended Irreversible Thermodynamics, Springer-Verlag, Berlin, 1993.
\bibitem{Jou_etal_1999} D. Jou, J. Casas-V\'azquez, G. Lebon,\ti{ Extended irreversible thermodynamics revisited (1988-98),} Rep. Prog. Phys. {\vol 62}, 1035\ti{-1142} (1999).
%
\bibitem{Crucifix_2007_pc} M. Crucifix, {\it pers. comm.}, 2007.
%
\bibitem{Prigogine_1980} I. Prigogine, From Being to Becoming: Time and Complexity in the Physical Sciences, W.H. Freeman \& Co., San Francisco, 1980.
\bibitem{Prigogine_S_1984} I. Prigogine, I. Stengers, Order Out of Chaos: Man's New Dialogue with Nature, Fontana Paperbacks, London, 1984.
\bibitem{Bejan_1997d}A. Bejan, Advanced Engineering Thermodynamics, 2nd ed., Wiley, NY, 1997, chap.\ 13.
\bibitem{Schneider_S_2005} E.D. Schneider, D. Sagan, Into the Cool: Energy Flow, Thermodynamics and Life, University of Chicago Press, Chicago, 2005.
\bibitem{Schrodinger_1944}E. Schr\"odinger, What is Life?, Cambridge U.P., Cambridge, 1944.
\bibitem{Kauffmann_2004} S.A. Kauffmann,\ti{ Autonomous agents,} {\it in} J.D. Barrow, P.C.W. Davies, C.L. Harper Jr. (eds) Science and Ultimate Reality: Quantum Theory, Cosmology and Complexity, Cambridge Univ. Press, 2004, 654\ti{-666}. 
%
\bibitem{Brillouin_1953} L. Brillouin,\ti{ The negentropy principle of information,} J. Appl. Phys. {\vol 24}(9), 1152\ti{-1163} (1953).
\bibitem{Curie_1894} P. Curie,\ti{ Sur la sym\'{e}trie dans les ph\'{e}nom\`{e}nes physiques, sym\'{e}trie d'un champ \'{e}lectrique et d'un champ magn\'{e}tique,} J. Physique Radium, {\vol 3}(3), 393\ti{-415} (1894); P. Curie, Oeuvres, Gauthier-Villars, Paris, 1908, 118\ti{-141}. Although widely cited, this work only provides a tenuous form of the ``Curie postulate''; see critique by Truesdell \cite{Truesdell_1969}.
\bibitem{Truesdell_1969} C. Truesdell, Rational Thermodynamics, McGraw-Hill, NY, 1969, chap.\ 7.

\end{thebibliography}
\end{document}